\documentclass[prl,twocolumn,superscriptaddress,showpacs,preprintnumbers,amsmath,amssymb]{revtex4-1}

\usepackage{graphicx}
\usepackage{dcolumn}
\usepackage{bm}
\usepackage{enumerate}
\usepackage{mathrsfs}
\usepackage{amsmath}
\usepackage{epstopdf}
\usepackage[titletoc]{appendix}
\usepackage[colorlinks=true,breaklinks=true,linkcolor=blue,citecolor=blue,urlcolor=blue]{hyperref}
\usepackage{color}

\newcommand{\Tr}{\rm Tr}
\begin{document}

\title{Uniaxial dynamical decoupling for an open quantum system}

\author{Qi Yao}
\affiliation{School of Physics and Technology, Wuhan University, Wuhan,
Hubei 430072, China}

\author{Jun Zhang}
\affiliation{School of Physics and Technology, Wuhan University, Wuhan,
Hubei 430072, China}

\author{Xiao-Feng Yi}
\affiliation{School of Physics and Technology, Wuhan University, Wuhan,
Hubei 430072, China}

\author{Li You}
\affiliation{State Key Laboratory of Low-Dimensional Quantum Physics, Department of Physics, Tsinghua University, Beijing 100084, China}

\author{Wenxian Zhang}
\email[Corresponding email: ]{wxzhang@whu.edu.cn}
\affiliation{School of Physics and Technology, Wuhan University, Wuhan, Hubei 430072, China}

\date{\today}

\begin{abstract}
Dynamical decoupling (DD) is an active and effective method for suppressing decoherence of a quantum system from its environment. In contrast to the nominal biaxial DD,
this work presents a uniaxial decoupling protocol that requires a significantly reduced number of pulses and a much lower bias field satisfying the ``magic" condition.
We show this uniaxial DD protocol works effectively in a number of model systems of practical interests, e.g., a spinor atomic Bose-Einstein condensate in stray magnetic fields (classical noise), or an electron spin coupled to nuclear spins (quantum noise) in a semiconductor quantum dot. It requires only half the number of control pulses and
a 10-100 times lower bias field for decoupling as normally employed in the above mentioned illustrative examples,
and the overall efficacy is robust against rotation errors of the control pulses.
The uniaxial DD protocol we propose shines new light on coherent controls in quantum computing and quantum information processing, quantum metrology, and low field nuclear magnetic resonance.
\end{abstract}

\maketitle

{\it Introduction.---}
Decoherence, due to coupling of a system to its surrounding environment, is a key obstacle towards practical applications of quantum technologies~\cite{Zurek1991Decoherence, nielsen2000quantum, Suter2016Colloquium}. Reliable quantum operations cannot proceed effectively or coherently without the decoherence of a quantum state under control~\cite{Knill2005Nature}. One may naively hope for the existence of a system perfectly isolated from its environment. However, this imposes heavy resource requirement and extreme conditions, such as ultralow temperature, ultrahigh vacuum, ultraweak/ultrastrong magnetic fields~\cite{pethick2002bose, Xiang2013Hybrid, Kelly2015State}, etc, some of which for all practical reasons cannot be achieved. Alternatively, one can search for strategies capable of slowing down or suppressing decoherence.

Dynamical decoupling (DD) is one such frequently employed decoherence-suppression method. It is capable of reducing effectively the coupling structure as well as the strength between a quantum system and its environment, thereby decoupling (or isolating) the system from its environment~\cite{Hahn1950Spin, haeberlen1976high, mehring1983principles, slichter1992principles}. DD has been widely employed for more than half a century in nuclear magnetic resonance (NMR) to isolate subsystems of nuclear spins from nearby spins and more recently in demonstrating robust quantum memory and universal quantum gate operations~\cite{viola1999Dynamical, Viola2003Robust, PhysRevLett.83.4888, Bacon2000Universal, PhysRevLett.102.080501, West2010High, SlavaNature2012}. Experiments in electron-nuclear spins and nitrogen vacancy centers have further established it as a powerful technique, e.g., capable of preserving arbitrary quantum states over extended times~\cite{Koppens2006Driven, SlavaNature2012, Medford2012Scaling, Amasha2008Electrical, Jelezko2004Obsevation, Xu2012Coherence}.

Most DD protocols employ biaxial resonant rotations, which in the presence of a large bias magnetic field along the $z$-axis (the quantization axis), can be constructed in terms of two rotations along two orthogonal (e.g., $x$- or $y$-) axes respectively. These resonant rotations can be either discrete or continuous, respectively associated with bang-bang DD or continuous DD~\cite{haeberlen1976high, viola1999Dynamical, viola2005Random, Kern2005controlling, Khodjasteh2005Fault, PhysRevA.71.022302, Santos2006enhanced, Witzel2007concatenated, Zhang2007dynamical, Uhrig2007keeping, yang2008Universality, Xu2012Coherence, Ball2015, Wang2017Delayed, Qi2017}. For bang-bang DD, many sophisticated protocols have been developed, e.g., periodic DD (PDD) or concatenated DD, which requires hard (delta function) pulses with a total number scaling as $4L$ for $L$ cycles and progression to even $4^L$ for $L$-level concatenation~\cite{Khodjasteh2005Fault, Zhang2008Long}. Several advanced schemes were utilized to optimize DD protocols, respectively given rise to the Uhrig DD~\cite{Uhrig2007keeping}, {concatenated Uhrig DD~\cite{Uhrig2009Concatenated}, and quadratic DD~\cite{West2010Near-Optimal},} with variable pulse delays to suppress high order noise correlations{~\cite{Wang2012Comparison}}. A question of great importance is: can one reduce the number of pulses, e.g., to $2L$ for $L$-cycle DD while maintaining the same level of noise suppression as in $4L$ PDD?

The uniaxial DD (Uni-DD) protocol we present in this Letter achieves such a challenging goal, making it a more efficient replacement for the usual biaxial {P}DD protocol. We show that the number of pulses reduces to the order of $2L$ for $L$-cycle DD while the performance remains similar to or better than the usual $4L$ PDD protocol. In addition, the $z$-axis bias magnetic field is reduced to about 100 times the average noise fields in the examples we studied, which is much less than the 1,000$\sim$10,000 times typically required in NMR experiments~\cite{mehring1983principles, Yannoni1976New, Zhang2009NMR}. Numerical simulations reveal the superior performance of our Uni-DD in a spinor Bose-Einstein condensate (BEC) decohered by stray magnetic fields and in a semiconductor quantum dot (QD) electron spin qubit decohered by nuclear spins. Our result can be applied to research in low field DD in quantum information, NMR, magnetic resonance imaging, and quantum sensing beyond standard quantum limit~\cite{Gross2008Squeezing, Degen2017Quantum, Luo2017Deterministic}.

{\it Uni-DD protocol.---}
We first briefly review the usual PDD protocol which forms the basis of the more advanced DD protocols. A qubit is decohered in general by stochastic interactions along three orthogonal directions, the longitudinal along the $z$-axis causes dephasing while the transversal along the $x$- or $y$-axis induces bit flip. Biaxial DD protocol suppresses the transversal noise by $\pi$ pulses along the $z$-axis (denoted as $Z$ pulses) and the longitudinal noise by $\pi$ pulses along either $x$- or $y$-axis (denoted as $X$ or $Y$ pulses, respectively)~\cite{viola1999Dynamical, Zhang2008Long}.

\begin{figure}
  \includegraphics[width=2in]{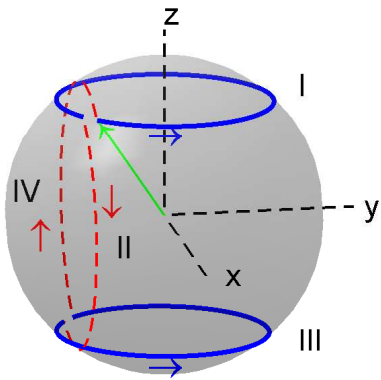}
  \caption{The schematic illustration for one cycle of the Uni-DD [$YU_\tau YU_\tau$] on the Bloch sphere. (I) A spin/qubit precesses around the total field composed of a large (longitudinal) bias and a small stochastic field for a duration $\tau$ (blue solid lines); (II) rotated by a hard $Y$ pulse (red dashed lines); (III) precesses around the total field for another $\tau$; (IV) and rotated by a second $Y$ pulse. The green solid line denotes the initial spin/qubit state.}
  \label{fig:sketch}
\end{figure}

The Uni-DD protocol we present also suppresses noise in all three directions. Similar to the biaxial DD, $Y$ pulses are employed to suppress dephasing from $z$-axis noise. Unlike the biaxial DD, a relatively strong bias magnetic field along $z$-axis is introduced to suppress transversal noises along $x$- or $y$-axis (see Fig.~\ref{fig:sketch}). The main inspiration to our idea comes from the observation that a strong longitudinal magnetic field suppresses transverse fluctuating fields~\cite{Shenvi2005Universal}. We further require the pulse delay $\tau$ and the effective Larmor precession frequency $\omega$ to satisfy $\omega \tau = n 2\pi$, with $n$ a positive integer. At this ``magic" condition, the qubit processes an integer number of rounds in the bias magnetic field between $Y$ pulses.

The Uni-DD protocol can be denoted in short hand as $[YU_{\tau} YU_{\tau}]^L$ for $L$-cycle with $U_{\tau}$ the precession operator, and the pulse delay $\tau$ satisfying the ``magic" condition within the shortest decoherence time possible. The latter is often given by the inhomogeneous broadening induced lifetime $T_2^*$~\cite{Merkulov2002Electron}. More details on the above results can be found in the supplemental material (SM), where we show the number of pulses for the Uni-DD is $2L$, or about half of the $4L$ pulses required by the PDD. In the following, we consider two concrete examples illustrating that our Uni-DD protocol is capable of suppressing classical or quantum noise.


\begin{figure}
  \includegraphics[width=3.25in]{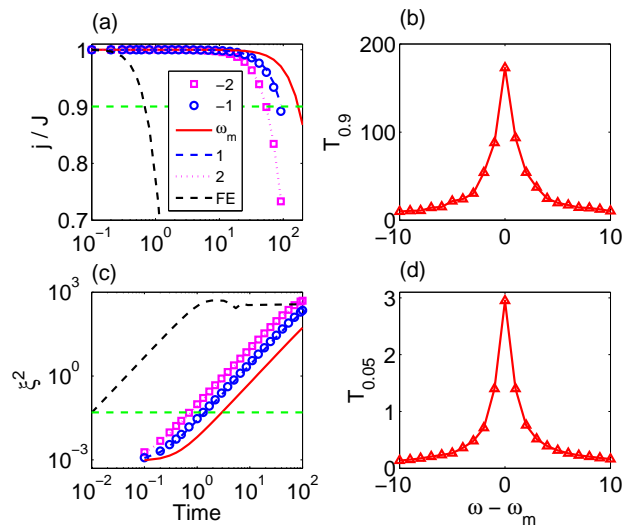}
  \caption{Suppressing classical noise with Uni-DD in a spinor atomic BEC for $\tau = 0.05$.
  (a) The evolution of the spin average under Uni-DD cycles for an initial CSS in the worst case at $\omega-\omega_m = -2$ (magenta squares), $-1$ (blue circles), $0$ (red solid line), $1$ (blue dashed line), and $2$ (magenta dotted line). The ``magic" condition requires $\omega_m = 2\pi / \tau \approx 126$. The FE results are presented for easier comparisons. The horizontal green dashed line denotes how the characteristic time $T_{0.9}$, where $j/J=0.9$, is extracted. (b) Dependence of the enhanced coherence time on the Larmor frequency (or the bias field). A peak occurs at the ``magic" condition $\omega_m \tau = 2\pi$. (c) Same as (a) except for the squeezing parameter $\xi^2$ with an initial SSS. The horizontal green dashed line is for $T_{0.05}$ where $\xi^2 = 0.05$. (d) Same as (b) except for $T_{0.05}$ with an initial SSS.}
  \label{fig:bec}
\end{figure}

{\it Suppressing classical stray magnetic fields in a spinor BEC.---}
As a model system decohered by classical noise, we consider a ferromagnetically interacting spin-1 atomic BEC
 under stray magnetic fields. Its full quantum state evolution is simulated including the Uni-DD protocol~\cite{zhang2016preserving}. Such a model allows the condensate spin degrees of freedom
 to be treated in terms of a large collective spin ${\mathbf J}$ (with $J=10^3$), which decohers by the stochastic rotations due to weak stray magnetic fields~\cite{Stamper2013Spinor, Law1998Quantum, yi2002single}. In the absence of the Uni-DD pulses, the model system is described by the Hamiltonian $H = c_2' {\mathbf J}^2 + \omega J_z + \gamma {\mathbf b}\cdot {\mathbf J}$, with $c_2'$ the effective atomic spin exchange interaction strength, $\omega=\gamma B$  the Larmor frequency in the longitudinal bias magnetic field $B$ ($\gamma$ is the gyro-magnetic ratio), and ${\mathbf b}$ the stray magnetic field stochastic in its direction and amplitude with a cutoff $b_c$ ($b_{x,y,z} \in [-b_c,b_c]$). For simplicity, the chosen stray field distribution function mimics a white noise, with the probability density for each realization uniformly distributed in direction and
amplitude over a limited range. We set $\hbar=1$, $\gamma = 1$, and $b_c=1$ for numerical simulations. The energy and the time units are $b_c$ and $b_c^{-1}$, respectively. For the Uni-DD, we choose a pulse delay of $\tau = 0.05$. The Uni-DD $Y$ pulses are assumed to be hard $\pi$ pulses, i.e., a temporal delta function with zero pulse width~\cite{Eto2014Control, parameternote}.

The simulations are carried out for four initial condensate spin directions, respectively along $x$-, $y$-, $z$-, and $-z$-axis, and the initial quantum state is either a coherent spin state (CSS) or a squeezed spin state (SSS). The worst performing case among the four is reported as a benchmark. We monitor the normalized spin average $j / J$ with $j = \sqrt{\langle {J_x} \rangle^2+\langle {J_y} \rangle^2+\langle {J_z} \rangle^2}$ for an initial CSS and the squeezing parameter $\xi^2 = 2{\rm min}\{\Delta J_x^2, \Delta J_y^2, \Delta J_z^2\} / J$ with $\Delta J_\alpha^2 = \langle J_\alpha^2 \rangle - \langle J_\alpha\rangle^2$ ($\alpha = x, y, z$) for an initial SSS during the Uni-DD~\cite{Kitagawa1993Squeezed, Morsch2006Dynamics, Jian2011Quantum}. In the absence of noises, the normalized spin average and the squeezing parameter should stay at unit and $0.00091$, respectively. Therefore, their respective rate of deterioration in the presence of noise manifests how fast the initial quantum system is decohered. Slower rate of deterioration due to the Uni-DD pulses than under pure free evolution (FE) indicates noise suppression.

The coherence time of the condensate spin in an initial CSS is clearly seen prolonged by two orders of magnitude as shown in Fig.~\ref{fig:bec}(a) and (b), if the ``magic" condition $\omega \tau = 2\pi$ is satisfied. For the more ``quantum" initial SSS, which is highly entangled and strongly correlated and thus expected to be more fragile or sensitive to noise, the coherence time is also seen prolonged by two orders of magnitude under the same ``magic" condition [Fig.~\ref{fig:bec}(c) and (d)]{~\cite{mcnote}}.

Enhanced understanding is gained by calculating analytically the evolution operator for a unit cycle of Uni-DD: $U_{2\tau} = [YU_{\tau}YU_{\tau}]$. Following the Fer expansion under the ``magic" condition $\omega\tau = 2\pi$, we find $U_{\tau}\approx \exp(-i\tau c_2'J^2) \exp(-i\tau {H}_{F,1})\exp(-i\tau {H}_{F,0}) $ with ${H}_{F,0} = \gamma b_z J_z$ and ${H}_{F,1} = \gamma^2 (b_z/\omega)(b_x J_x + b_y J_y) + J_z (b_x^2+b_y^2)/(2\omega)$~\cite{Fer1958R, takegoshi2015comparison}. By further employing the Magnus expansion, we obtain $U_{2\tau} \approx \exp(-i2\tau c_2'J^2)\exp[-i2\tau \gamma^2 (b_z/\omega)b_yJ_y]$ to the leading nonzero order~\cite{haeberlen1976high, mehring1983principles,
takegoshi2015comparison}. The derivation details can be found in the SM. Compared to the corresponding FE operator $U_{FE} =\exp(-i2\tau c_2'J^2)\exp[-i2\tau \gamma (b_xJ_x+b_yJ_y+b_zJ_z)]$ without the bias magnetic field, the effective coupling strength between the condensate spin and the stray magnetic field is seen to be reduced by a factor of $b_{z}/B$, which can become much smaller. Thus, noise suppression of the Uni-DD protocol is rooted in the bias field's suppression of the transversal fluctuation field and augmented by the cancelation of the longitudinal fluctuation field from the $Y$ pulses. This is quite different from the nominal biaxial DD protocol, which often relies on the smallness of the pulse delay.

{\it Suppressing nuclear spin quantum noise in a QD.---}
Unlike the classical environment described by stochastic complex fields, a proper description for a quantum environment must deal with environment operators and their correlations. To illustrate the power of the Uni-DD, we choose a gate-defined GaAs semiconductor QD system which is well described by a central spin model with the electron spin decohered by the surrounding nuclear spins~\cite{Loss1998Quantum, Petta2005Coherent, Koppens2006Driven, Taylor2007Relaxation, Zhang2007dynamical}. To further simplify the problem, we assume the electron spin (${\bf S}$) as well as all nuclear spins (${\bf I}_k$) are spin-1/2. The coupling $A_k$ between ${\mathbf S}$ and ${\mathbf I}_k$ results from the Fermi contact hyperfine interaction~\cite{paget1977low, Merkulov2002Electron}. The Hamiltonian for the model system of $N$ nuclear spins without Uni-DD takes the form $H = {\mathbf S}\cdot \sum_{k=1}^N A_k {\mathbf I}_k+\sum_{i< j}^N \Gamma_{ij}({\mathbf I}_i\cdot{\mathbf I}_j-3{\mathbf I}_{iz}\cdot{\mathbf I}_{jz})$. In general, $A_k$ is proportional to the local density of the electron at the position of the $k$th nucleon. In this work, it is modeled as in Ref.~\cite{dobrovitski2006long} for $N=4\times 5$ nuclear spins, by $A_k \propto \exp[-(x-x_0)^2/w_x^2-(y-y_0)^2/w_y^2]$, a 2D Gaussian form with effective widths $w_x/a_x = 3/2$ and $w_y/a_y=2$ and a shifted center $x_0/a_x=0.1$ and $y_0/a_y=0.2$, leading to the final values of $A_k$ ranging between 0.309 and 0.960. $\Gamma_{ij}$ accounts for the magnetic dipolar interaction of nearest neighbor nuclear spins and is randomly distributed between 0 and 0.01.

\begin{figure}
  \includegraphics[width=3.25in]{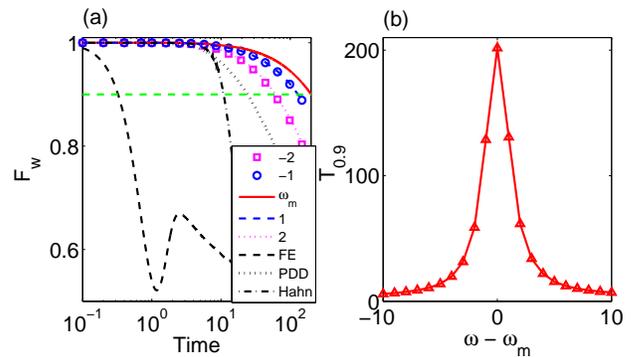}
  \caption{Suppression of quantum noise by the Uni-DD in a GaAs QD for $\tau = 0.05$. (a) The evolution of fidelity under Uni-DD cycles in the worst case for $\omega-\omega_m = -2$ (magenta squares), $-1$ (blue circles), $0$ (red solid line), $1$ (blue dashed line), and $2$ (magenta dotted line). The FE (black dashed line), Hahn echo (black dash-dotted line), and PDD (black dotted line) results are also presented for comparisons. The horizontal green dashed line denotes how the characteristic time $T_{0.9}$ is extracted. (b) Dependence of the prolonged coherence time on the Larmor frequency (or the bias field), exhibiting a peak at the ``magic" condition $\omega_m \tau = 2\pi$. }\label{fig:qd}
\end{figure}

It is well known for a GaAs QD system that the short time FE quantum dynamics agrees well with the classical (quasi-static bath approximation) fluctuation model, but the long time dynamics gradually deviates, implicating the important role played by quantum correlations within the coupled central spin system~\cite{zhang2006hyperfine1, Merkulov2002Electron, coish2008exponential}. We expect our Uni-DD would prolong the coherent dynamics,
and quantum simulations are thus carried out to fully account for the quantum corrections.

Starting from the electron spin initially pointing along $x$-, $y$-, $z$-, and $-z$-axis respectively, the system evolution is simulated quantum mechanically with the Chebyshev polynomial expansion method~\cite{dobrovitski2003efficient}. The initial nuclear spin state is a fully mixed state which is approximated numerically by a random pure state~\cite{Popescu2006Entanglement, nielsen2000quantum}. The evolution is recorded,
and the fidelity, which calibrates the survival probability of the initial electron spin state, $F(t) = \Tr[\rho_e(0) \Tr_n(\rho(t))]$, is calculated, where $\Tr_n(\rho(t))$ is the reduced electron spin state after tracing out all nuclear spins. The worst case fidelity, $F_w = \min_{\{\rho_e(0)\}} (F)$, among the four initial states
is easily identified and used as a benchmark and shown in Fig.~\ref{fig:qd}. Compared to FE, the coherence time of the electron spin is prolonged by upto two orders of magnitude with Uni-DD, depending on whether the condition $\omega\tau = 2\pi$ is satisfied or not [Fig.~\ref{fig:qd}(a)]. Compared to PDD, the Uni-DD is also found to be superior. The enhanced coherence time implicates successful decoupling of the electron spin from its surrounding nuclear spins. More interestingly, the ``magic" condition exhibits a resonance,
around which the characteristic coherence time of the Uni-DD protocol $T_{0.9}$ versus the Larmor frequency $\omega$ of the bias field ($\omega = \gamma B$) is shown in Fig.~\ref{fig:qd}(b).

The decoupling by the Uni-DD protocol under the ``magic" condition for the QD model can again be proven analytically by following the Fer expansion of $U_\tau$ with the average Hamiltonian theory based on the Magnus expansion. Although the average Hamiltonian theory does not directly apply since the convergence condition $|H|\tau \ll 1$ is violated, its application becomes possible after adopting the Fer expansion of $U_{\tau}$ in a rotating reference frame defined by the bias field~\cite{Fer1958R, takegoshi2015comparison}. In fact, Fer expansion is applicable at treating long time quantum evolution beyond the convergence radius of the widely used Magus expansion~\cite{haeberlen1976high, mehring1983principles, takegoshi2015comparison}. After a straightforward derivation at the ``magic" condition $\omega\tau = 2\pi$, we find $U_{\tau} \approx \exp[-i\tau H_{F,1}] \exp[-i\tau H_{F,0}]$ where $H_{F,0} = S_z h_z$ and $H_{F,1} = S_x(h_zh_x+h_xh_z)/(2\omega)$$+S_y(h_zh_y+h_yh_z)/(2\omega)$$ +S_z(h_x^2+h_y^2)/(2\omega)$$ +i(h_xh_y-h_yh_x)/(4\omega)$ with the quantum Overhauser field operator $h_{\alpha\in\{x,y,z\}} =\sum_{k=1}^N A_k I_{k\alpha}$. For one Uni-DD cycle, the evolution operator reduces to $U_{2\tau} = [Y U_{\tau} Y U_{\tau}] \approx \exp(-i2\tau[S_y(h_zh_y+h_yh_z)/(2\omega)+i(h_xh_y-h_yh_x)/(4\omega)])$, as shown in detail in the SM. Similar to the classical noise example considered earlier, one sees immediately that the relatively strong bias field suppresses the relaxation effect of the transversal quantum noise and the $Y$ pulses suppress the dephasing effect of the longitudinal quantum noise~{\cite{biasnote}}.

\begin{figure}
  \includegraphics[width=3.2in]{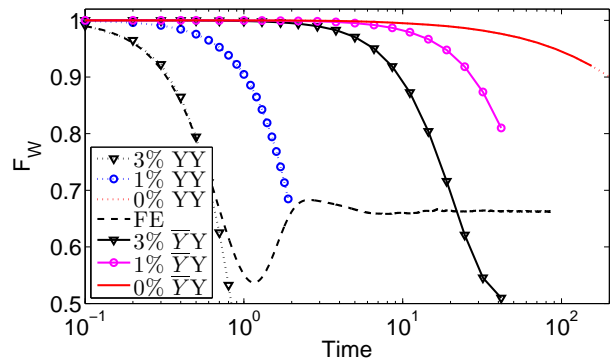}
  \caption{Robustness of the Uni-DD against rotation angle errors for $[YU_\tau YU_\tau]$ with $\varepsilon = 3\%$ (black dotted line with triangles), $1\%$ (blue dotted line with circles), $0\%$ (red dotted line) and for $[\overline{Y}U_\tau YU_\tau]$ with $\varepsilon = 3\%$ (black solid line with triangles), $1\%$ (blue solid line with circles), $0\%$ (red solid line, coinciding with the red dotted line). The FE results are also presented for comparisons.}\label{fig:err}
\end{figure}

{\it Robustness of the Uni-DD against rotation angle errors.---}
To estimate the robustness of the Uni-DD protocol, we consider rotation angle error $\varepsilon$ of the $Y$ pulses, i.e., assuming an imperfect rotation angle of the $Y$ pulse $(1-\varepsilon) \pi$. As shown in Fig.~\ref{fig:err}, even a small $\varepsilon = 1\%$ causes the worst case fidelity $F_w$ to drop significantly. Such a sensitive dependence on the rotation angle can be remedied by replacing one of the $Y$ pulses with a $\overline{Y}$ pulse, which rotates along the $-y$ direction with the same imperfect angle $(1-\varepsilon) \pi$. This idea is behind the Carr-Purcell-Meiboom-Gill (CPMG) protocol which improves greatly the robustness of the Carr-Purcell protocol~\cite{slichter1992principles, Carr1954Effects, Meiboom1958Modified}. Remarkably, the modified Uni-DD protocol likewise shows strong robustness against the rotation angle error, as illustrated in Fig.~\ref{fig:err}. Even for $\varepsilon = 3\%$, the coherence time remains prolonged by an order of magnitude. For a smaller $\varepsilon = 1\%$, the Uni-DD protocol is seen to almost reach the same outcome as in the perfect pulse case of $\varepsilon = 0$.

Finally, we note the Uni-DD protocol we discuss differs fundamentally from the single pulse Hahn echo by requiring the ``magic" condition. Such a condition in the Uni-DD allows the bias field to stay as low as possible while keeping the noise suppression effect comparable or even better than in the Hahn echo, as shown in Fig.~\ref{fig:qd}(a) (and more in the SM). {By Adopting symmetrization and concatenation, more advanced DD protocols based on Uni-DD may be developed with improved performances. Our preliminary investigations into the performance comparisons of the Uni-DD with the standard symmetrized DD, the second level concatenated DD, the nonequidistant concatenated Uhrig DD, and the quadratic DD are presented in the SM. More efforts could be devoted to explore systematically these protocols in order to find the most suitable one for a specific experiment~\cite{nvcnote}.}

In conclusion, we propose a Uni-DD protocol for suppressing the decoherence of an open quantum system from its environment. Compared to the nominal biaxial PDD, the Uni-DD achieves the same degree of noise suppression with half the number of control pulses. We demonstrate with numerical and analytical calculations the efficacy of the Uni-DD under the ``magic" condition $\omega \tau = 2\pi$ in suppressing the classical stray fields in a spinor BEC and in suppressing the quantum nuclear spin noises in a GaAs QD. Our results point to an alternative low-cost DD techniques which may find wide applications in quantum computing and quantum information processing, NMR and magnetic resonance imaging, as well as quantum precision measurements beyond the standard quantum limit.

\begin{acknowledgments}
W.Z. thanks Dr. V. V. Dobrovitski for inspiring discussions. This work is supported by the National Natural Science Foundation of China (NSFC) under Grant Nos. 11574239 and 91836101 and by the Open Research Fund Program of the State Key Laboratory of Low Dimensional Quantum Physics under Grant No. KF201614. L.Y. acknowledges the support by the National Key R$\&$D Program of China (Grant No. 2018YFA0306504) and by NSFC (Grant No. 11654001).
\end{acknowledgments}

\begin{appendices}
\vbox{}
\centerline{\textbf{Supplementary Materials}}

\section{I. Suppressing classical noise with Uni-DD}

For the collective spin of an atomic condensate $\mathbf{J}$ in the presence of a bias magnetic field along the $z$-axis and classical stray magnetic fields, the Hamiltonian of the system takes the form~\cite{Ho1998Spinor, Ohmi1998Bose,
Law1998Quantum1, Zhang2003Mean1, zhang2016preserving1}
\begin{equation}\label{eq:hbec}
  H = c_2' {\mathbf J}^2 + \omega J_z + \gamma (b_z J_z + b_x J_x + b_y J_y).
\end{equation}
In this study we take $J = 1000$. $c_2'<0$ is the spin exchange interaction strength in a ferromagnetic spin-1 atomic Bose-Einstein condensate (BEC) under the single spatial mode approximation, $\omega = \gamma B$ with the bias magnetic field $B$ and the gyro-magnetic ratio $\gamma$, and ${\mathbf b}$ the stray magnetic fields. In general, the bias field is much stronger than the stray magnetic field $B \gg |{\mathbf b}|$. The effect of the stray magnetic field is simulated by averaging $r=100$ random realizations of $b_{x,y,z} \in [-b_c, b_c]$ with $b_c=1$ the cutoff magnitude. The distributions of $b_{x,y,z}$ are independent and identical resembling white noise.

The decoherence of the condensate spin is caused by the stray magnetic field during free evolution (FE) in a zero bias field. Without the bias field or dynamic decoupling (DD) pulses, their coupling strength is of the order of $b_c$. The spin exchange interaction term $\propto{\mathbf J}^2$ does not contribute to decoherence. The application
of Uni-DD protocol reduces the effective coupling strength. In the following, we show how much is this reduction.

When the Uni-DD is applied, the evolution of condensate spin in a stray magnetic field is governed by
\begin{equation}\label{eq:u2tau}
  U_{2\tau} = YU_{\tau}YU_{\tau} \approx \exp(-i2\tau \overline H),
\end{equation}
where $\overline H$ is the effective Hamiltonian, $U_{\tau} = \exp(-i\tau H)$, and $Y$ the $\pi$ pulse along $y$-axis. We separate approximately the evolution of the spin in the bias field from that in the stray field
in the following manner
\begin{equation}\label{eq:utau}
  U_{\tau} = \exp(-i\tau c_2'{\mathbf J}^2) R {\cal T}: \exp(-i\int_0^\tau H_R(t') dt'),
\end{equation}
where the rotating operator $R = \exp(-i\omega t J_z)$ defines a rotating reference frame in which the time-dependent Hamiltonian is
\begin{eqnarray}\label{eq:hr}
  H_R(t) = &\gamma & b_z J_z + \gamma b_x[J_x\cos(\omega t)-J_y\sin(\omega t)] + \nonumber \\
   &\gamma & b_y[J_y\cos(\omega t)+J_x\sin(\omega t)].
\end{eqnarray}

By employing the Fer expansion and the ``magic" condition $\omega \tau = 2\pi$, we
find after a lengthy calculation~\cite{Fer1958R1, takegoshi2015comparison1}
\begin{equation}
  {\cal T}: \exp(-i\int_0^\tau H_R(t') dt') \approx  \exp(-i\tau H_{F,1}) \exp(-i\tau H_{F,0}),
\end{equation}
to the second order $(b_{x,y,z})^2/B$ with $H_{F,1} =  \gamma (b_z/B)(b_x J_x + b_y J_y) + \gamma J_z (b_x^2+b_y^2)/(2B)$ and $H_{F,0} = \gamma b_z J_z$. Because $R=I$ under the ``magic" condition,
we obtain the following evolution operator
\begin{equation}\label{eq:utau2}
  U_{\tau} \approx \exp(-i\tau c_2'{\mathbf J}^2) \exp(-i\tau H_{F,1}) \exp(-i\tau H_{F,0}).
\end{equation}
Since $\gamma b_{x,y,z}\tau \ll 1$ and $b_c \ll B$, it is
appropriate to adopt the average Hamiltonian theory based on the Magnus
expansion to calculate for the whole cycle the effective Hamiltonian given by~\cite{Haeberlen1968Coherent, haeberlen1976high, mehring1983principles,
takegoshi2015comparison1}
\begin{equation}\label{eq:hbar}
  \overline H = c_2'{\mathbf J}^2 + \frac{b_z}B \; \gamma b_y J_y.
\end{equation}
Compared to the free Hamiltonian Eq.~(\ref{eq:hbec}), the effective Hamiltonian $\overline H$ shows that the coupling strength between the condensate spin and the stray magnetic field is reduced by a small factor $b_z/B$, which explains why the Uni-DD is effective in suppressing stray magnetic fields. Additionally, the suppression effect is shown to be independent of the initial state for the condensate spin, as shown in Fig.~2 in the main text.

\section{II. Suppressing quantum noise with Uni-DD}

For an electron spin ${\mathbf S}$ interacting with surrounding nuclear spins
${\mathbf I}_k$ in a QD at a bias magnetic field $B$ with a Larmor frequency $\omega = \gamma B$, the free Hamiltonian takes the form~\cite{Loss1998Quantum1, Petta2005Coherent1, Koppens2006Driven1, Taylor2007Relaxation1, al2006numerical, zhang2006hyperfine1, zhang2007dynamical1}
\begin{equation}\label{eq:hqd}
  H = \omega S_z + h_x S_x + h_y S_y + h_z S_z+\sum_{i<j}^N \Gamma_{ij}({\mathbf I}_i\cdot{\mathbf I}_j-3{\mathbf I}_i^z\cdot{\mathbf I}_j^z),
\end{equation}
where the quantum noise operators are $h_{x,y,z} = \sum_{k=1}^N A_k I_{k\{x,y,z\}}$ with $A_k$ the coupling constant. The last term in Eq.~(\ref{eq:hqd}) describing only correlations of quantum noises is neglected because it commutes with the central electron spin. Different from the classical noise field $b_{x,y,z}$, the quantum noise operators $h_{x,y,z}$ in general do not commute.

Following the same procedure as outlined in the above for classical noise, we define a rotating reference frame with respect to the bias field $B$ in which the time-dependent Hamiltonian becomes
\begin{eqnarray}
  H_R(t) = & h_z & S_z + h_x[S_x\cos(\omega t)-S_y\sin(\omega t)]+ \nonumber \\
  & h_y & [S_y\cos(\omega t)+S_x\sin(\omega t)].
\end{eqnarray}
Under the same ``magic" condition $\omega \tau = 2\pi$, the evolution operator $U_{\tau}$ becomes
\begin{eqnarray}\label{eq:utauqd}
  U_{\tau} &=& R {\cal T}:\exp(-i\int_0^\tau H_R(t') dt') \nonumber \\
    &\approx & \exp(-i\tau H_{F,1})\exp(-i\tau H_{F,0}),
\end{eqnarray}
where $R=I$, $H_{F,1}=S_x({h_zh_x+h_xh_z})/{2\omega} +S_y({h_zh_y+ h_yh_z})/{2\omega}+S_z({h_x^2+h_y^2})/{2\omega}
+i({h_xh_y-h_yh_x})/{4\omega}$, and $H_{F,0} = h_z S_z$, and the Fer expansion (valid at long times~\cite{Fer1958R1,
takegoshi2015comparison1}) is applied. The above result reduces exactly to the
classical one if we ignore the non-commuting nature of $h_{x,y,z}$.

By further employing the average Hamiltonian
theory~\cite{Haeberlen1968Coherent, haeberlen1976high,
mehring1983principles, takegoshi2015comparison1}, it is straightforward to find the
effective Hamiltonian for one Uni-DD cycle
\begin{equation}\label{eq:hbarqd}
  \overline H = S_y({h_zh_y+h_yh_z})/{2\omega} + i({h_xh_y-h_yh_x})/{4\omega}.
\end{equation}
Compared to the free Hamiltonian Eq.~(\ref{eq:hqd}) of the QD system, the effective coupling constant in the average Hamiltonian is reduced also by a small factor $|h_z|/\omega$. The last term contains only nuclear spin operators thus only affects the electron spin dynamics at long times. The suppression effect on the quantum noises is also independent of the initial electron spin state.

\section{III. Comparison of the Uni-DD with the Hahn echo protocol}

As shown in Fig.~3(a) in the main text, the Uni-DD protocol outperforms the traditional Hahn echo in suppressing the electron spin's decoherence by the nuclear spins in a quantum dot. While in a spin-1 atomic BEC, the conclusion
remains the same, either for an initial coherent spin state (CSS) or a squeezed spin state (SSS).

To systematically compare the Uni-DD with the Hahn echo, we vary the correlation time $\tau_c$ of the stray magnetic fields, from a value as small as 0.5 to a much larger value for a total evolution time $t=100$. For each correlation time $\tau_c$, we average the results over 100 simulations. During each run of simulation, we fix the correlation time $\tau_c$, i.e., changing the random magnetic field ${\mathbf b}_n$ to a new random value ${\mathbf b}_{n+1}$ at the evolution time $t= (n+1)\tau_c$. To make a ``fair" comparison, we also take the ``magic" condition for the Hahn echo $YU({t/ 2})YU({t/ 2})$, i.e., $\omega (t/ 2) = k\cdot 2\pi$, where $U({t/ 2})$ is the free evolution and $k$ an integer.

\begin{figure}
  \includegraphics[width=3.25in]{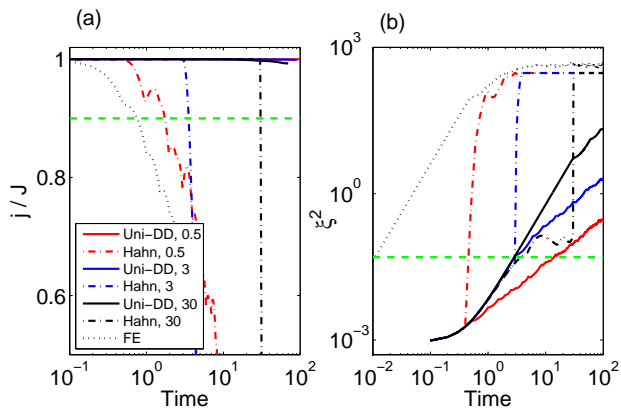}
  \caption{Comparison of the performances for the Uni-DD (solid lines) and Hahn echo (dash-dotted lines) protocol in a spin-1 atomic BEC at $\tau_c = 0.5$ (red lines), $3$ (blue lines), and $30$ (black lines). The pulse delay of the Uni-DD is $\tau = 0.05$ and the ``magic" condition $\omega \tau = 2\pi$ is satisfied. The worst-case spin average $j/J$ for an initial CSS is present in (a) and the worst-case squeezing parameter $\xi^2$ for an initial SSS in (b). The free evolution results (dotted lines) are also shown in (a) and (b). The Uni-DD is seen to clearly outperform the Hahn echo in suppressing decoherence by the stray magnetic field, particularly at short noise correlation times.}
  \label{fig:sm1}
\end{figure}

Typical results for $\tau_c = 0.5$, $3$, and $30$ are shown in Fig.~\ref{fig:sm1}(a) for an initial CSS and in Fig.~\ref{fig:sm1}(b) for an initial SSS. As in the main text, the worst case among the four chosen initial states
is reported. In Fig.~\ref{fig:sm1}(a) we find the spin average deviates from one much slower under the Uni-DD than under the Hahn echo, indicating that the Uni-DD outperforms. Interestingly, the spin average under the Uni-DD protocol decreases gradually after the evolution time $t>\tau_c$, while the spin average under the Hahn echo decreases abruptly after $t>\tau_c$. Similarly, we find in Fig.~\ref{fig:sm1}(b) that the squeezing parameter increases slowly after $t>\tau_c$ for the Uni-DD but jumps up abruptly for the Hahn echo. In all, the Uni-DD is seen to also outperform the Hahn echo for an initial SSS.

\section{IV. Preliminary comparisons of uniaxial and biaxial DD protocols}

By adopting the methods of symmetrization and concatenation, symmetrized Uni-DD (SUni-DD) or concatenated Uni-DD (CUni-DD) protocols can be designed. For example, a simple SUni-DD cycle is $[\overline{U}_\tau Y\overline{U}_\tau U_\tau YU_\tau]$ where $\overline{U}_\tau$ denotes free evolution for an interval $\tau$ but with a reversed bias magnetic field, i.e., $\omega = -\gamma B$. It is easy to check the first order terms proportional to $\omega^{-1}$ are canceled out cleanly due to symmetrization. We also introduce a simple concatenation by
setting SUni-DD as the first level concatenation $C_1$ = SUni-DD, and taking the second level concatenation
as $C_2 = [\overline{C}_1 Y\overline{C}_1 C_1 YC_1]$.

\begin{figure}
  \includegraphics[width=3.25in]{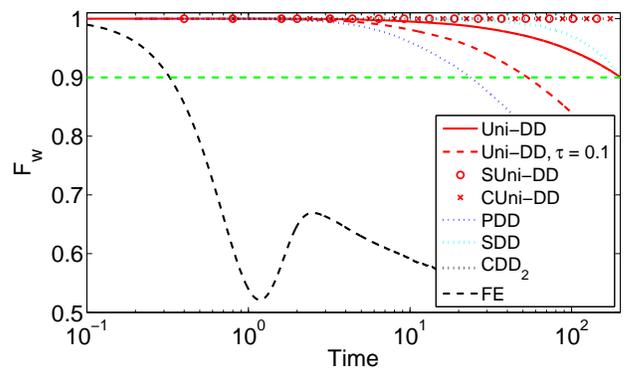}
  \caption{Comparison of the performances for several equidistant DD protocols, including Uni-DD (red solid line), Uni-DD at $\tau = 0.1$ (red dashed line), SUni-DD (red circles), CUni-DD$_2$ (crosses), PDD (blue dotted line), SDD (cyan dotted line), CDD$_2$ (black dotted line), and FE (black dashed line). The parameters used are
$\tau = 0.05$ (unless otherwise noted) and satisfying the ``magic" condition for the uniaxial DD protocols. The horizontal green dashed line marks the 90\% fidelity.}
  \label{fig:sm2}
\end{figure}

Figure~\ref{fig:sm2} compares the performances of the SUni-DD and the CUni-DD$_2$ protocol.
Unlike the Uni-DD, the fidelities of the SUni-DD and the CUni-DD$_2$ are seen to
always stay close to 1. Due to the power of compensating for the higher order terms
from symmetrization and concatenation, the corresponding coherence times are significantly extended. We also present in Fig.~\ref{fig:sm2} the Uni-DD result at $\tau = 0.1$ with the ``magic" condition enforced, i.e., keeping $\omega = 2\pi/\tau = 62.8$. Although the fidelity decays faster than at $\tau = 0.05$,
the Uni-DD with $\tau = 0.1$ still extends the coherence time about 2 orders of magnitude.

The PDD protocol can also be designed as a symmetrized DD (SDD) or a second-level concatenated DD (CDD$_2$)~\cite{khodjasteh2005Fault, zhang2007suppression, Zhang2008Long1}. The corresponding pulse sequences are $[ZU_{\tau}XU_{\tau}ZU_{\tau}XU_{\tau}]$ for the PDD, $[U_{\tau}XU_{\tau}ZU_{\tau}XU_{\tau} U_{\tau}XU_{\tau}ZU_{\tau}XU_{\tau}]$ for the SDD, and $[ZC_1 XC_1 ZC_1 XC_1]$ for the CDD$_2$ with $C_1$=PDD. The simulation results under these standard biaxial DD protocols are presented in Fig.~\ref{fig:sm2}. Clearly, the SDD performance is better than the PDD but worse than the CDD$_2$, since the CDD$_2$ is capable of
nulling out higher order terms than the SDD and the PDD. In addition, the fidelity hardly decays under the CDD$_2$ pulse sequence, and is almost indistinguishable from the SUni-DD and the CUni-DD in the region computed.

\begin{figure}
  \includegraphics[width=3.25in]{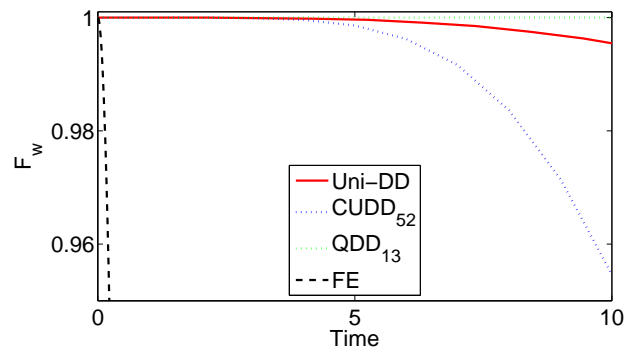}
  \caption{Comparison of the performances for the Uni-DD (red solid lines) with nonequidistant biaxial sequences CUDD$_{52}$ (blue dotted line) and QDD$_{13}$ (black dotted line), for the same pulse number $N_p=210$. The free evolution results (black dashed line) are also presented. The horizontal green dashed line marks the 90\% fidelity.}
  \label{fig:sm3}
\end{figure}

An alternative biaxial DD family is based on Uhrig's nonequidistant DD, including the concatenated Uhrig DD (CUDD) and the quadratic DD (QDD). The CUDD for $n$-th order cancelation during a total evolution time $t$ is~\cite{Uhrig2007keeping1, Uhrig2009Concatenated1, Wang2012Comparison1},
$CUDD_n= U_n^{(Z)}(t/4)XU_n^{(Z)}(t/4)U_n^{(Z)}(t/4)XU_n^{(Z)}(t/4)$ with the inner concatenation sequence $U_n^{(Z)}(t)= U_{t-t_n}ZU_{\tau_n}Z \cdot\cdot\cdot U_{\tau_2}ZU_{\tau_1}$, where the interval $\tau_j=t_j-t_{j-1}$ is determined by a Uhrig sequence $t_j=t\sin^2(\frac{j\pi}{2N_p-2})$ with $N_p = 4n+2$ the total pulse number. The QDD with an odd $n$-th order cancelation during a total evolution time $t$ is~\cite{West2010Near-Optimal1}
$QDD_n = XU_n^{(Z)}(\delta_{n+1}) X U_n^{(Z)}(\delta_n) X \cdot\cdot\cdot X U_n^{(Z)}(\delta_2) X U_n^{(Z)}(\delta_1)$ with the inner concatenation sequences $U_n^{(Z)}(t)= ZU_{t-t_n}ZU_{\tau_n}Z \cdot\cdot\cdot U_{\tau_2}ZU_{\tau_1}$ and the interval $\delta_j=\varepsilon_j-\varepsilon_{j-1}$ which is also determined by the Uhrig sequences $\varepsilon_j=t\sin^2(\frac{j\pi}{2N_p-2})$. The total number of pulses is $N_p = (n+1)(n+2)$.

To make comparisons among CUDD, QDD, and Uni-DD, we fix the number of applied pulses $N_p = 210$ and change the total evolution time $t$ by varying the interval $\tau$ (and $\omega = 2 \pi/\tau$ for the Uni-DD). Other parameters are the same as the quantum dot model in the main text, except that the dipolar couplings between environmental nuclear spins are set five times larger, in order to illustrate the difference among these protocols. Figure~\ref{fig:sm3} shows there comparisons. Clearly, all DD protocols are seen to extend significantly the coherence time. The QDD$_{13}$ outperforms the Uni-DD and the CUDD$_{52}$, while the performance of the Uni-DD lies in between, although it
only compensates for the lowest order terms. By exploiting symmetrization and concatenation, we expect much improved performance for all uniaxial DD protocols such as the SUni-DD and CUni-DD.

\section{V. Suppression of spin noises in nitrogen vacancy center with Uni-DD protocol}

\begin{figure}
  \includegraphics[width=3.25in]{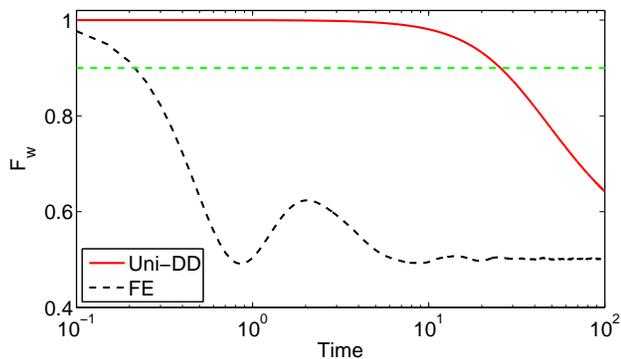}
  \caption{Performance for the Uni-DD (red solid line) in a NV center. The worst-case fidelity $F_w(t)$ under Uni-DD protocol is graphed as a function of time. The result for free evolution (black dashed line) is also presented as a comparison. The horizontal green dashed line marks the 90\% fidelity. The coherence time is prolonged about 2 orders of magnitude.}
  \label{fig:sm4}
\end{figure}

In the above, the Uni-DD protocol is shown to extend the coherence time of a quantum system with isotropic (Heisenberg type) coupling to the environment. This section illustrates that it is also effective for other types of coupling, e.g., the dipolar coupling between a central spin and its environment composed of other spins. A well-known example is a nitrogen vacancy (NV) center in a diamond, where an important coupling between the NV center spin and the surrounding nuclear spins is~\cite{Hall2014PRB}
$$ H= \sum_{k=1}^{N} A_k\left({\mathbf S_0} \cdot {\mathbf S_k} -3S_0^z S_k^z\right),$$
with ${\mathbf S_0}$ denoting the NV center spin, ${\mathbf S_k}$ the $k$th nuclear spin, and $A_k$ the dipolar coupling strength. We assume all spins are spin-1/2 and the coupling $A_k$ is distributed randomly between 0 and 1. Such a dipolar coupling is anisotropic.

Figure~\ref{fig:sm4} shows the numerical results we obtained. The decoherence time of the free evolution is about $T_{0.9} \approx 0.2$, which is prolonged by the Uni-DD pulse sequence to $T_{0.9} \approx 25$.
Again, an enhancement of 2 orders of magnitude is observed.

 \end{appendices}


\begin{thebibliography}{71}%
\makeatletter
\providecommand \@ifxundefined [1]{%
 \@ifx{#1\undefined}
}%
\providecommand \@ifnum [1]{%
 \ifnum #1\expandafter \@firstoftwo
 \else \expandafter \@secondoftwo
 \fi
}%
\providecommand \@ifx [1]{%
 \ifx #1\expandafter \@firstoftwo
 \else \expandafter \@secondoftwo
 \fi
}%
\providecommand \natexlab [1]{#1}%
\providecommand \enquote  [1]{``#1''}%
\providecommand \bibnamefont  [1]{#1}%
\providecommand \bibfnamefont [1]{#1}%
\providecommand \citenamefont [1]{#1}%
\providecommand \href@noop [0]{\@secondoftwo}%
\providecommand \href [0]{\begingroup \@sanitize@url \@href}%
\providecommand \@href[1]{\@@startlink{#1}\@@href}%
\providecommand \@@href[1]{\endgroup#1\@@endlink}%
\providecommand \@sanitize@url [0]{\catcode `\\12\catcode `\$12\catcode
  `\&12\catcode `\#12\catcode `\^12\catcode `\_12\catcode `\%12\relax}%
\providecommand \@@startlink[1]{}%
\providecommand \@@endlink[0]{}%
\providecommand \url  [0]{\begingroup\@sanitize@url \@url }%
\providecommand \@url [1]{\endgroup\@href {#1}{\urlprefix }}%
\providecommand \urlprefix  [0]{URL }%
\providecommand \Eprint [0]{\href }%
\providecommand \doibase [0]{http://dx.doi.org/}%
\providecommand \selectlanguage [0]{\@gobble}%
\providecommand \bibinfo  [0]{\@secondoftwo}%
\providecommand \bibfield  [0]{\@secondoftwo}%
\providecommand \translation [1]{[#1]}%
\providecommand \BibitemOpen [0]{}%
\providecommand \bibitemStop [0]{}%
\providecommand \bibitemNoStop [0]{.\EOS\space}%
\providecommand \EOS [0]{\spacefactor3000\relax}%
\providecommand \BibitemShut  [1]{\csname bibitem#1\endcsname}%
\let\auto@bib@innerbib\@empty
\bibitem [{\citenamefont {Zurek}(1991)}]{Zurek1991Decoherence}%
  \BibitemOpen
  \bibfield  {author} {\bibinfo {author} {\bibfnamefont {W.~H.}\ \bibnamefont
  {Zurek}},\ }\href {https://doi.org/10.1063/1.881293} {\bibfield  {journal}
  {\bibinfo  {journal} {Phys. Today}\ }\textbf {\bibinfo {volume} {44}},\
  \bibinfo {pages} {36} (\bibinfo {year} {1991})}\BibitemShut {NoStop}%
\bibitem [{\citenamefont {Nielsen}\ and\ \citenamefont
  {Chuang}(2000)}]{nielsen2000quantum}%
  \BibitemOpen
  \bibfield  {author} {\bibinfo {author} {\bibfnamefont {M.~A.}\ \bibnamefont
  {Nielsen}}\ and\ \bibinfo {author} {\bibfnamefont {I.~L.}\ \bibnamefont
  {Chuang}},\ }\href@noop {} {\emph {\bibinfo {title} {Quantum information and
  computation}}}\ (\bibinfo  {publisher} {Cambridge University Press,
  Cambridge},\ \bibinfo {year} {2000})\BibitemShut {NoStop}%
\bibitem [{\citenamefont {Suter}\ and\ \citenamefont
  {\'Alvarez}(2016)}]{Suter2016Colloquium}%
  \BibitemOpen
  \bibfield  {author} {\bibinfo {author} {\bibfnamefont {D.}~\bibnamefont
  {Suter}}\ and\ \bibinfo {author} {\bibfnamefont {G.~A.}\ \bibnamefont
  {\'Alvarez}},\ }\href {\doibase 10.1103/RevModPhys.88.041001} {\bibfield
  {journal} {\bibinfo  {journal} {Rev. Mod. Phys.}\ }\textbf {\bibinfo {volume}
  {88}},\ \bibinfo {pages} {041001} (\bibinfo {year} {2016})}\BibitemShut
  {NoStop}%
\bibitem [{\citenamefont {Knill}(2005)}]{Knill2005Nature}%
  \BibitemOpen
  \bibfield  {author} {\bibinfo {author} {\bibfnamefont {E.}~\bibnamefont
  {Knill}},\ }\href {https://doi.org/10.1038/nature03350} {\bibfield  {journal}
  {\bibinfo  {journal} {Nature (London)}\ }\textbf {\bibinfo {volume} {434}},\
  \bibinfo {pages} {39} (\bibinfo {year} {2005})}\BibitemShut {NoStop}%
\bibitem [{\citenamefont {Pethick}\ and\ \citenamefont
  {Smith}(2002)}]{pethick2002bose}%
  \BibitemOpen
  \bibfield  {author} {\bibinfo {author} {\bibfnamefont {C.~J.}\ \bibnamefont
  {Pethick}}\ and\ \bibinfo {author} {\bibfnamefont {H.}~\bibnamefont
  {Smith}},\ }\href@noop {} {\emph {\bibinfo {title} {Bose-Einstein
  condensation in dilute gases}}}\ (\bibinfo  {publisher} {Cambridge University
  Press, New York},\ \bibinfo {year} {2002})\BibitemShut {NoStop}%
\bibitem [{\citenamefont {Xiang}\ \emph {et~al.}(2013)\citenamefont {Xiang},
  \citenamefont {Ashhab}, \citenamefont {You},\ and\ \citenamefont
  {Nori}}]{Xiang2013Hybrid}%
  \BibitemOpen
  \bibfield  {author} {\bibinfo {author} {\bibfnamefont {Z.-L.}\ \bibnamefont
  {Xiang}}, \bibinfo {author} {\bibfnamefont {S.}~\bibnamefont {Ashhab}},
  \bibinfo {author} {\bibfnamefont {J.~Q.}\ \bibnamefont {You}}, \ and\
  \bibinfo {author} {\bibfnamefont {F.}~\bibnamefont {Nori}},\ }\href {\doibase
  10.1103/RevModPhys.85.623} {\bibfield  {journal} {\bibinfo  {journal} {Rev.
  Mod. Phys.}\ }\textbf {\bibinfo {volume} {85}},\ \bibinfo {pages} {623}
  (\bibinfo {year} {2013})}\BibitemShut {NoStop}%
\bibitem [{\citenamefont {Kelly}\ \emph {et~al.}(2015)\citenamefont {Kelly},
  \citenamefont {Barends}, \citenamefont {Fowler}, \citenamefont {Megrant},
  \citenamefont {Jeffrey}, \citenamefont {White}, \citenamefont {Sank},
  \citenamefont {Mutus}, \citenamefont {Campbell},\ and\ \citenamefont
  {Chen}}]{Kelly2015State}%
  \BibitemOpen
  \bibfield  {author} {\bibinfo {author} {\bibfnamefont {J.}~\bibnamefont
  {Kelly}}, \bibinfo {author} {\bibfnamefont {R.}~\bibnamefont {Barends}},
  \bibinfo {author} {\bibfnamefont {A.~G.}\ \bibnamefont {Fowler}}, \bibinfo
  {author} {\bibfnamefont {A.}~\bibnamefont {Megrant}}, \bibinfo {author}
  {\bibfnamefont {E.}~\bibnamefont {Jeffrey}}, \bibinfo {author} {\bibfnamefont
  {T.~C.}\ \bibnamefont {White}}, \bibinfo {author} {\bibfnamefont
  {D.}~\bibnamefont {Sank}}, \bibinfo {author} {\bibfnamefont {J.~Y.}\
  \bibnamefont {Mutus}}, \bibinfo {author} {\bibfnamefont {B.}~\bibnamefont
  {Campbell}}, \ and\ \bibinfo {author} {\bibfnamefont {Y.}~\bibnamefont
  {Chen}},\ }\href {\doibase 10.1038/npjqi.2015.11} {\bibfield  {journal}
  {\bibinfo  {journal} {Nature (London)}\ }\textbf {\bibinfo {volume} {519}},\
  \bibinfo {pages} {66} (\bibinfo {year} {2015})}\BibitemShut {NoStop}%
\bibitem [{\citenamefont {Hahn}(1950)}]{Hahn1950Spin}%
  \BibitemOpen
  \bibfield  {author} {\bibinfo {author} {\bibfnamefont {E.~L.}\ \bibnamefont
  {Hahn}},\ }\href {\doibase 10.1103/PhysRev.80.580} {\bibfield  {journal}
  {\bibinfo  {journal} {Phys. Rev.}\ }\textbf {\bibinfo {volume} {80}},\
  \bibinfo {pages} {580} (\bibinfo {year} {1950})}\BibitemShut {NoStop}%
\bibitem [{\citenamefont {Haeberlen}(1976)}]{haeberlen1976high}%
  \BibitemOpen
  \bibfield  {author} {\bibinfo {author} {\bibfnamefont {U.}~\bibnamefont
  {Haeberlen}},\ }\href@noop {} {\emph {\bibinfo {title} {High Resolution NMR
  in solids: selective averaging}}}\ (\bibinfo  {publisher} {Academic, New
  York},\ \bibinfo {year} {1976})\BibitemShut {NoStop}%
\bibitem [{\citenamefont {Mehring}(1983)}]{mehring1983principles}%
  \BibitemOpen
  \bibfield  {author} {\bibinfo {author} {\bibfnamefont {M.}~\bibnamefont
  {Mehring}},\ }\href@noop {} {\emph {\bibinfo {title} {Principles of high
  resolution NMR in solids}}},\ \bibinfo {edition} {2nd}\ ed.\ (\bibinfo
  {publisher} {Springer-Verleg, Berlin},\ \bibinfo {year} {1983})\BibitemShut
  {NoStop}%
\bibitem [{\citenamefont {Slichter}(1992)}]{slichter1992principles}%
  \BibitemOpen
  \bibfield  {author} {\bibinfo {author} {\bibfnamefont {C.~P.}\ \bibnamefont
  {Slichter}},\ }\href@noop {} {\emph {\bibinfo {title} {Principles of Magnetic
  Resonance}}}\ (\bibinfo  {publisher} {Springer-Verlag, New York},\ \bibinfo
  {year} {1992})\BibitemShut {NoStop}%
\bibitem [{\citenamefont {Viola}\ \emph
  {et~al.}(1999{\natexlab{a}})\citenamefont {Viola}, \citenamefont {Knill},\
  and\ \citenamefont {Lloyd}}]{viola1999Dynamical}%
  \BibitemOpen
  \bibfield  {author} {\bibinfo {author} {\bibfnamefont {L.}~\bibnamefont
  {Viola}}, \bibinfo {author} {\bibfnamefont {E.}~\bibnamefont {Knill}}, \ and\
  \bibinfo {author} {\bibfnamefont {S.}~\bibnamefont {Lloyd}},\ }\href
  {\doibase 10.1103/PhysRevLett.82.2417} {\bibfield  {journal} {\bibinfo
  {journal} {Phys. Rev. Lett.}\ }\textbf {\bibinfo {volume} {82}},\ \bibinfo
  {pages} {2417} (\bibinfo {year} {1999}{\natexlab{a}})}\BibitemShut {NoStop}%
\bibitem [{\citenamefont {Viola}\ and\ \citenamefont
  {Knill}(2003)}]{Viola2003Robust}%
  \BibitemOpen
  \bibfield  {author} {\bibinfo {author} {\bibfnamefont {L.}~\bibnamefont
  {Viola}}\ and\ \bibinfo {author} {\bibfnamefont {E.}~\bibnamefont {Knill}},\
  }\href {\doibase 10.1103/PhysRevLett.90.037901} {\bibfield  {journal}
  {\bibinfo  {journal} {Phys. Rev. Lett.}\ }\textbf {\bibinfo {volume} {90}},\
  \bibinfo {pages} {037901} (\bibinfo {year} {2003})}\BibitemShut {NoStop}%
\bibitem [{\citenamefont {Viola}\ \emph
  {et~al.}(1999{\natexlab{b}})\citenamefont {Viola}, \citenamefont {Lloyd},\
  and\ \citenamefont {Knill}}]{PhysRevLett.83.4888}%
  \BibitemOpen
  \bibfield  {author} {\bibinfo {author} {\bibfnamefont {L.}~\bibnamefont
  {Viola}}, \bibinfo {author} {\bibfnamefont {S.}~\bibnamefont {Lloyd}}, \ and\
  \bibinfo {author} {\bibfnamefont {E.}~\bibnamefont {Knill}},\ }\href
  {\doibase 10.1103/PhysRevLett.83.4888} {\bibfield  {journal} {\bibinfo
  {journal} {Phys. Rev. Lett.}\ }\textbf {\bibinfo {volume} {83}},\ \bibinfo
  {pages} {4888} (\bibinfo {year} {1999}{\natexlab{b}})}\BibitemShut {NoStop}%
\bibitem [{\citenamefont {Bacon}\ \emph {et~al.}(2000)\citenamefont {Bacon},
  \citenamefont {Kempe}, \citenamefont {Lidar},\ and\ \citenamefont
  {Whaley}}]{Bacon2000Universal}%
  \BibitemOpen
  \bibfield  {author} {\bibinfo {author} {\bibfnamefont {D.}~\bibnamefont
  {Bacon}}, \bibinfo {author} {\bibfnamefont {J.}~\bibnamefont {Kempe}},
  \bibinfo {author} {\bibfnamefont {D.~A.}\ \bibnamefont {Lidar}}, \ and\
  \bibinfo {author} {\bibfnamefont {K.~B.}\ \bibnamefont {Whaley}},\ }\href
  {\doibase 10.1103/PhysRevLett.85.1758} {\bibfield  {journal} {\bibinfo
  {journal} {Phys. Rev. Lett.}\ }\textbf {\bibinfo {volume} {85}},\ \bibinfo
  {pages} {1758} (\bibinfo {year} {2000})}\BibitemShut {NoStop}%
\bibitem [{\citenamefont {Khodjasteh}\ and\ \citenamefont
  {Viola}(2009)}]{PhysRevLett.102.080501}%
  \BibitemOpen
  \bibfield  {author} {\bibinfo {author} {\bibfnamefont {K.}~\bibnamefont
  {Khodjasteh}}\ and\ \bibinfo {author} {\bibfnamefont {L.}~\bibnamefont
  {Viola}},\ }\href {\doibase 10.1103/PhysRevLett.102.080501} {\bibfield
  {journal} {\bibinfo  {journal} {Phys. Rev. Lett.}\ }\textbf {\bibinfo
  {volume} {102}},\ \bibinfo {pages} {080501} (\bibinfo {year}
  {2009})}\BibitemShut {NoStop}%
\bibitem [{\citenamefont {West}\ \emph
  {et~al.}(2010{\natexlab{a}})\citenamefont {West}, \citenamefont {Lidar},
  \citenamefont {Fong},\ and\ \citenamefont {Gyure}}]{West2010High}%
  \BibitemOpen
  \bibfield  {author} {\bibinfo {author} {\bibfnamefont {J.~R.}\ \bibnamefont
  {West}}, \bibinfo {author} {\bibfnamefont {D.~A.}\ \bibnamefont {Lidar}},
  \bibinfo {author} {\bibfnamefont {B.~H.}\ \bibnamefont {Fong}}, \ and\
  \bibinfo {author} {\bibfnamefont {M.~F.}\ \bibnamefont {Gyure}},\ }\href
  {\doibase 10.1103/PhysRevLett.105.230503} {\bibfield  {journal} {\bibinfo
  {journal} {Phys. Rev. Lett.}\ }\textbf {\bibinfo {volume} {105}},\ \bibinfo
  {pages} {230503} (\bibinfo {year} {2010}{\natexlab{a}})}\BibitemShut
  {NoStop}%
\bibitem [{\citenamefont {van~der Sar}\ \emph {et~al.}(2012)\citenamefont
  {van~der Sar}, \citenamefont {Wang}, \citenamefont {Blok}, \citenamefont
  {Bernien}, \citenamefont {Taminiau}, \citenamefont {Toyli}, \citenamefont
  {Lidar}, \citenamefont {Awschalom}, \citenamefont {Hanson},\ and\
  \citenamefont {Dobrovitski}}]{SlavaNature2012}%
  \BibitemOpen
  \bibfield  {author} {\bibinfo {author} {\bibfnamefont {T.}~\bibnamefont
  {van~der Sar}}, \bibinfo {author} {\bibfnamefont {Z.~H.}\ \bibnamefont
  {Wang}}, \bibinfo {author} {\bibfnamefont {M.~S.}\ \bibnamefont {Blok}},
  \bibinfo {author} {\bibfnamefont {H.}~\bibnamefont {Bernien}}, \bibinfo
  {author} {\bibfnamefont {T.~H.}\ \bibnamefont {Taminiau}}, \bibinfo {author}
  {\bibfnamefont {D.~M.}\ \bibnamefont {Toyli}}, \bibinfo {author}
  {\bibfnamefont {D.~A.}\ \bibnamefont {Lidar}}, \bibinfo {author}
  {\bibfnamefont {D.~D.}\ \bibnamefont {Awschalom}}, \bibinfo {author}
  {\bibfnamefont {R.}~\bibnamefont {Hanson}}, \ and\ \bibinfo {author}
  {\bibfnamefont {V.~V.}\ \bibnamefont {Dobrovitski}},\ }\href {\doibase
  10.1038/nature10900} {\bibfield  {journal} {\bibinfo  {journal} {Nature
  (London)}\ }\textbf {\bibinfo {volume} {484}},\ \bibinfo {pages} {82}
  (\bibinfo {year} {2012})}\BibitemShut {NoStop}%
\bibitem [{\citenamefont {Koppens}\ \emph {et~al.}(2006)\citenamefont
  {Koppens}, \citenamefont {Buizert}, \citenamefont {Tielrooij}, \citenamefont
  {Vink}, \citenamefont {Nowack}, \citenamefont {Meunier}, \citenamefont
  {Kouwenhoven},\ and\ \citenamefont {Vandersypen}}]{Koppens2006Driven}%
  \BibitemOpen
  \bibfield  {author} {\bibinfo {author} {\bibfnamefont {F.~H.}\ \bibnamefont
  {Koppens}}, \bibinfo {author} {\bibfnamefont {C.}~\bibnamefont {Buizert}},
  \bibinfo {author} {\bibfnamefont {K.~J.}\ \bibnamefont {Tielrooij}}, \bibinfo
  {author} {\bibfnamefont {I.~T.}\ \bibnamefont {Vink}}, \bibinfo {author}
  {\bibfnamefont {K.~C.}\ \bibnamefont {Nowack}}, \bibinfo {author}
  {\bibfnamefont {T.}~\bibnamefont {Meunier}}, \bibinfo {author} {\bibfnamefont
  {L.~P.}\ \bibnamefont {Kouwenhoven}}, \ and\ \bibinfo {author} {\bibfnamefont
  {L.~M.}\ \bibnamefont {Vandersypen}},\ }\href
  {https://doi.org/10.1038/nature05065} {\bibfield  {journal} {\bibinfo
  {journal} {Nature (London)}\ }\textbf {\bibinfo {volume} {442}},\ \bibinfo
  {pages} {766} (\bibinfo {year} {2006})}\BibitemShut {NoStop}%
\bibitem [{\citenamefont {Medford}\ \emph {et~al.}(2012)\citenamefont
  {Medford}, \citenamefont {Cywi\ifmmode~\acute{n}\else \'{n}\fi{}ski},
  \citenamefont {Barthel}, \citenamefont {Marcus}, \citenamefont {Hanson},\
  and\ \citenamefont {Gossard}}]{Medford2012Scaling}%
  \BibitemOpen
  \bibfield  {author} {\bibinfo {author} {\bibfnamefont {J.}~\bibnamefont
  {Medford}}, \bibinfo {author} {\bibfnamefont {L.}~\bibnamefont
  {Cywi\ifmmode~\acute{n}\else \'{n}\fi{}ski}}, \bibinfo {author}
  {\bibfnamefont {C.}~\bibnamefont {Barthel}}, \bibinfo {author} {\bibfnamefont
  {C.~M.}\ \bibnamefont {Marcus}}, \bibinfo {author} {\bibfnamefont {M.~P.}\
  \bibnamefont {Hanson}}, \ and\ \bibinfo {author} {\bibfnamefont {A.~C.}\
  \bibnamefont {Gossard}},\ }\href {\doibase 10.1103/PhysRevLett.108.086802}
  {\bibfield  {journal} {\bibinfo  {journal} {Phys. Rev. Lett.}\ }\textbf
  {\bibinfo {volume} {108}},\ \bibinfo {pages} {086802} (\bibinfo {year}
  {2012})}\BibitemShut {NoStop}%
\bibitem [{\citenamefont {Amasha}\ \emph {et~al.}(2008)\citenamefont {Amasha},
  \citenamefont {MacLean}, \citenamefont {Radu}, \citenamefont {Zumb\"uhl},
  \citenamefont {Kastner}, \citenamefont {Hanson},\ and\ \citenamefont
  {Gossard}}]{Amasha2008Electrical}%
  \BibitemOpen
  \bibfield  {author} {\bibinfo {author} {\bibfnamefont {S.}~\bibnamefont
  {Amasha}}, \bibinfo {author} {\bibfnamefont {K.}~\bibnamefont {MacLean}},
  \bibinfo {author} {\bibfnamefont {I.~P.}\ \bibnamefont {Radu}}, \bibinfo
  {author} {\bibfnamefont {D.~M.}\ \bibnamefont {Zumb\"uhl}}, \bibinfo {author}
  {\bibfnamefont {M.~A.}\ \bibnamefont {Kastner}}, \bibinfo {author}
  {\bibfnamefont {M.~P.}\ \bibnamefont {Hanson}}, \ and\ \bibinfo {author}
  {\bibfnamefont {A.~C.}\ \bibnamefont {Gossard}},\ }\href {\doibase
  10.1103/PhysRevLett.100.046803} {\bibfield  {journal} {\bibinfo  {journal}
  {Phys. Rev. Lett.}\ }\textbf {\bibinfo {volume} {100}},\ \bibinfo {pages}
  {046803} (\bibinfo {year} {2008})}\BibitemShut {NoStop}%
\bibitem [{\citenamefont {Jelezko}\ \emph {et~al.}(2004)\citenamefont
  {Jelezko}, \citenamefont {Gaebel}, \citenamefont {Popa}, \citenamefont
  {Gruber},\ and\ \citenamefont {Wrachtrup}}]{Jelezko2004Obsevation}%
  \BibitemOpen
  \bibfield  {author} {\bibinfo {author} {\bibfnamefont {F.}~\bibnamefont
  {Jelezko}}, \bibinfo {author} {\bibfnamefont {T.}~\bibnamefont {Gaebel}},
  \bibinfo {author} {\bibfnamefont {I.}~\bibnamefont {Popa}}, \bibinfo {author}
  {\bibfnamefont {A.}~\bibnamefont {Gruber}}, \ and\ \bibinfo {author}
  {\bibfnamefont {J.}~\bibnamefont {Wrachtrup}},\ }\href {\doibase
  10.1103/PhysRevLett.92.076401} {\bibfield  {journal} {\bibinfo  {journal}
  {Phys. Rev. Lett.}\ }\textbf {\bibinfo {volume} {92}},\ \bibinfo {pages}
  {076401} (\bibinfo {year} {2004})}\BibitemShut {NoStop}%
\bibitem [{\citenamefont {Xu}\ \emph {et~al.}(2012)\citenamefont {Xu},
  \citenamefont {Wang}, \citenamefont {Duan}, \citenamefont {Huang},
  \citenamefont {Wang}, \citenamefont {Wang}, \citenamefont {Xu}, \citenamefont
  {Kong}, \citenamefont {Shi}, \citenamefont {Rong},\ and\ \citenamefont
  {Du}}]{Xu2012Coherence}%
  \BibitemOpen
  \bibfield  {author} {\bibinfo {author} {\bibfnamefont {X.}~\bibnamefont
  {Xu}}, \bibinfo {author} {\bibfnamefont {Z.}~\bibnamefont {Wang}}, \bibinfo
  {author} {\bibfnamefont {C.}~\bibnamefont {Duan}}, \bibinfo {author}
  {\bibfnamefont {P.}~\bibnamefont {Huang}}, \bibinfo {author} {\bibfnamefont
  {P.}~\bibnamefont {Wang}}, \bibinfo {author} {\bibfnamefont {Y.}~\bibnamefont
  {Wang}}, \bibinfo {author} {\bibfnamefont {N.}~\bibnamefont {Xu}}, \bibinfo
  {author} {\bibfnamefont {X.}~\bibnamefont {Kong}}, \bibinfo {author}
  {\bibfnamefont {F.}~\bibnamefont {Shi}}, \bibinfo {author} {\bibfnamefont
  {X.}~\bibnamefont {Rong}}, \ and\ \bibinfo {author} {\bibfnamefont
  {J.}~\bibnamefont {Du}},\ }\href {\doibase 10.1103/PhysRevLett.109.070502}
  {\bibfield  {journal} {\bibinfo  {journal} {Phys. Rev. Lett.}\ }\textbf
  {\bibinfo {volume} {109}},\ \bibinfo {pages} {070502} (\bibinfo {year}
  {2012})}\BibitemShut {NoStop}%
\bibitem [{\citenamefont {Viola}\ and\ \citenamefont
  {Knill}(2005)}]{viola2005Random}%
  \BibitemOpen
  \bibfield  {author} {\bibinfo {author} {\bibfnamefont {L.}~\bibnamefont
  {Viola}}\ and\ \bibinfo {author} {\bibfnamefont {E.}~\bibnamefont {Knill}},\
  }\href {\doibase 10.1103/PhysRevLett.94.060502} {\bibfield  {journal}
  {\bibinfo  {journal} {Phys. Rev. Lett.}\ }\textbf {\bibinfo {volume} {94}},\
  \bibinfo {pages} {060502} (\bibinfo {year} {2005})}\BibitemShut {NoStop}%
\bibitem [{\citenamefont {Kern}\ and\ \citenamefont
  {Alber}(2005)}]{Kern2005controlling}%
  \BibitemOpen
  \bibfield  {author} {\bibinfo {author} {\bibfnamefont {O.}~\bibnamefont
  {Kern}}\ and\ \bibinfo {author} {\bibfnamefont {G.}~\bibnamefont {Alber}},\
  }\href {\doibase 10.1103/PhysRevLett.95.250501} {\bibfield  {journal}
  {\bibinfo  {journal} {Phys. Rev. Lett.}\ }\textbf {\bibinfo {volume} {95}},\
  \bibinfo {pages} {250501} (\bibinfo {year} {2005})}\BibitemShut {NoStop}%
\bibitem [{\citenamefont {Khodjasteh}\ and\ \citenamefont
  {Lidar}(2005)}]{Khodjasteh2005Fault}%
  \BibitemOpen
  \bibfield  {author} {\bibinfo {author} {\bibfnamefont {K.}~\bibnamefont
  {Khodjasteh}}\ and\ \bibinfo {author} {\bibfnamefont {D.~A.}\ \bibnamefont
  {Lidar}},\ }\href {\doibase 10.1103/PhysRevLett.95.180501} {\bibfield
  {journal} {\bibinfo  {journal} {Phys. Rev. Lett.}\ }\textbf {\bibinfo
  {volume} {95}},\ \bibinfo {pages} {180501} (\bibinfo {year}
  {2005})}\BibitemShut {NoStop}%
\bibitem [{\citenamefont {Facchi}\ \emph {et~al.}(2005)\citenamefont {Facchi},
  \citenamefont {Tasaki}, \citenamefont {Pascazio}, \citenamefont {Nakazato},
  \citenamefont {Tokuse},\ and\ \citenamefont {Lidar}}]{PhysRevA.71.022302}%
  \BibitemOpen
  \bibfield  {author} {\bibinfo {author} {\bibfnamefont {P.}~\bibnamefont
  {Facchi}}, \bibinfo {author} {\bibfnamefont {S.}~\bibnamefont {Tasaki}},
  \bibinfo {author} {\bibfnamefont {S.}~\bibnamefont {Pascazio}}, \bibinfo
  {author} {\bibfnamefont {H.}~\bibnamefont {Nakazato}}, \bibinfo {author}
  {\bibfnamefont {A.}~\bibnamefont {Tokuse}}, \ and\ \bibinfo {author}
  {\bibfnamefont {D.~A.}\ \bibnamefont {Lidar}},\ }\href {\doibase
  10.1103/PhysRevA.71.022302} {\bibfield  {journal} {\bibinfo  {journal} {Phys.
  Rev. A}\ }\textbf {\bibinfo {volume} {71}},\ \bibinfo {pages} {022302}
  (\bibinfo {year} {2005})}\BibitemShut {NoStop}%
\bibitem [{\citenamefont {Santos}\ and\ \citenamefont
  {Viola}(2006)}]{Santos2006enhanced}%
  \BibitemOpen
  \bibfield  {author} {\bibinfo {author} {\bibfnamefont {L.~F.}\ \bibnamefont
  {Santos}}\ and\ \bibinfo {author} {\bibfnamefont {L.}~\bibnamefont {Viola}},\
  }\href {\doibase 10.1103/PhysRevLett.97.150501} {\bibfield  {journal}
  {\bibinfo  {journal} {Phys. Rev. Lett.}\ }\textbf {\bibinfo {volume} {97}},\
  \bibinfo {pages} {150501} (\bibinfo {year} {2006})}\BibitemShut {NoStop}%
\bibitem [{\citenamefont {Witzel}\ and\ \citenamefont
  {Das~Sarma}(2007)}]{Witzel2007concatenated}%
  \BibitemOpen
  \bibfield  {author} {\bibinfo {author} {\bibfnamefont {W.~M.}\ \bibnamefont
  {Witzel}}\ and\ \bibinfo {author} {\bibfnamefont {S.}~\bibnamefont
  {Das~Sarma}},\ }\href {\doibase 10.1103/PhysRevB.76.241303} {\bibfield
  {journal} {\bibinfo  {journal} {Phys. Rev. B}\ }\textbf {\bibinfo {volume}
  {76}},\ \bibinfo {pages} {241303} (\bibinfo {year} {2007})}\BibitemShut
  {NoStop}%
\bibitem [{\citenamefont {Zhang}\ \emph {et~al.}(2007)\citenamefont {Zhang},
  \citenamefont {Dobrovitski}, \citenamefont {Santos}, \citenamefont {Viola},\
  and\ \citenamefont {Harmon}}]{Zhang2007dynamical}%
  \BibitemOpen
  \bibfield  {author} {\bibinfo {author} {\bibfnamefont {W.}~\bibnamefont
  {Zhang}}, \bibinfo {author} {\bibfnamefont {V.~V.}\ \bibnamefont
  {Dobrovitski}}, \bibinfo {author} {\bibfnamefont {L.~F.}\ \bibnamefont
  {Santos}}, \bibinfo {author} {\bibfnamefont {L.}~\bibnamefont {Viola}}, \
  and\ \bibinfo {author} {\bibfnamefont {B.~N.}\ \bibnamefont {Harmon}},\
  }\href {\doibase 10.1103/PhysRevB.75.201302} {\bibfield  {journal} {\bibinfo
  {journal} {Phys. Rev. B}\ }\textbf {\bibinfo {volume} {75}},\ \bibinfo
  {pages} {201302} (\bibinfo {year} {2007})}\BibitemShut {NoStop}%
\bibitem [{\citenamefont {Uhrig}(2007)}]{Uhrig2007keeping}%
  \BibitemOpen
  \bibfield  {author} {\bibinfo {author} {\bibfnamefont {G.~S.}\ \bibnamefont
  {Uhrig}},\ }\href {\doibase 10.1103/PhysRevLett.98.100504} {\bibfield
  {journal} {\bibinfo  {journal} {Phys. Rev. Lett.}\ }\textbf {\bibinfo
  {volume} {98}},\ \bibinfo {pages} {100504} (\bibinfo {year}
  {2007})}\BibitemShut {NoStop}%
\bibitem [{\citenamefont {Yang}\ and\ \citenamefont
  {Liu}(2008)}]{yang2008Universality}%
  \BibitemOpen
  \bibfield  {author} {\bibinfo {author} {\bibfnamefont {W.}~\bibnamefont
  {Yang}}\ and\ \bibinfo {author} {\bibfnamefont {R.-B.}\ \bibnamefont {Liu}},\
  }\href {\doibase 10.1103/PhysRevLett.101.180403} {\bibfield  {journal}
  {\bibinfo  {journal} {Phys. Rev. Lett.}\ }\textbf {\bibinfo {volume} {101}},\
  \bibinfo {pages} {180403} (\bibinfo {year} {2008})}\BibitemShut {NoStop}%
\bibitem [{\citenamefont {Ball}\ and\ \citenamefont
  {Biercuk}(2015)}]{Ball2015}%
  \BibitemOpen
  \bibfield  {author} {\bibinfo {author} {\bibfnamefont {H.}~\bibnamefont
  {Ball}}\ and\ \bibinfo {author} {\bibfnamefont {M.~J.}\ \bibnamefont
  {Biercuk}},\ }\href {\doibase 10.1140/epjqt/s40507-015-0022-4} {\bibfield
  {journal} {\bibinfo  {journal} {EPJ Quantum. Techol.}\ }\textbf {\bibinfo
  {volume} {2}},\ \bibinfo {pages} {11} (\bibinfo {year} {2015})}\BibitemShut
  {NoStop}%
\bibitem [{\citenamefont {Wang}\ \emph {et~al.}(2017)\citenamefont {Wang},
  \citenamefont {Casanova},\ and\ \citenamefont {Plenio}}]{Wang2017Delayed}%
  \BibitemOpen
  \bibfield  {author} {\bibinfo {author} {\bibfnamefont {Z.~Y.}\ \bibnamefont
  {Wang}}, \bibinfo {author} {\bibfnamefont {J.}~\bibnamefont {Casanova}}, \
  and\ \bibinfo {author} {\bibfnamefont {M.~B.}\ \bibnamefont {Plenio}},\
  }\href {https://doi.org/10.1038/ncomms14660} {\bibfield  {journal} {\bibinfo
  {journal} {Nat. Commun.}\ }\textbf {\bibinfo {volume} {8}},\ \bibinfo {pages}
  {14660} (\bibinfo {year} {2017})}\BibitemShut {NoStop}%
\bibitem [{\citenamefont {Qi}\ \emph {et~al.}(2017)\citenamefont {Qi},
  \citenamefont {Dowling},\ and\ \citenamefont {Viola}}]{Qi2017}%
  \BibitemOpen
  \bibfield  {author} {\bibinfo {author} {\bibfnamefont {H.}~\bibnamefont
  {Qi}}, \bibinfo {author} {\bibfnamefont {J.~P.}\ \bibnamefont {Dowling}}, \
  and\ \bibinfo {author} {\bibfnamefont {L.}~\bibnamefont {Viola}},\ }\href
  {\doibase 10.1007/s11128-017-1719-3} {\bibfield  {journal} {\bibinfo
  {journal} {Quantum Information Processing}\ }\textbf {\bibinfo {volume}
  {16}},\ \bibinfo {pages} {272} (\bibinfo {year} {2017})}\BibitemShut
  {NoStop}%
\bibitem [{\citenamefont {Zhang}\ \emph {et~al.}(2008)\citenamefont {Zhang},
  \citenamefont {Konstantinidis}, \citenamefont {Dobrovitski}, \citenamefont
  {Harmon}, \citenamefont {Santos},\ and\ \citenamefont
  {Viola}}]{Zhang2008Long}%
  \BibitemOpen
  \bibfield  {author} {\bibinfo {author} {\bibfnamefont {W.}~\bibnamefont
  {Zhang}}, \bibinfo {author} {\bibfnamefont {N.~P.}\ \bibnamefont
  {Konstantinidis}}, \bibinfo {author} {\bibfnamefont {V.~V.}\ \bibnamefont
  {Dobrovitski}}, \bibinfo {author} {\bibfnamefont {B.~N.}\ \bibnamefont
  {Harmon}}, \bibinfo {author} {\bibfnamefont {L.~F.}\ \bibnamefont {Santos}},
  \ and\ \bibinfo {author} {\bibfnamefont {L.}~\bibnamefont {Viola}},\ }\href
  {\doibase 10.1103/PhysRevB.77.125336} {\bibfield  {journal} {\bibinfo
  {journal} {Phys. Rev. B.}\ }\textbf {\bibinfo {volume} {77}},\ \bibinfo
  {pages} {125336} (\bibinfo {year} {2008})}\BibitemShut {NoStop}%
\bibitem [{\citenamefont {Uhrig}(2009)}]{Uhrig2009Concatenated}%
  \BibitemOpen
  \bibfield  {author} {\bibinfo {author} {\bibfnamefont {G.~S.}\ \bibnamefont
  {Uhrig}},\ }\href {\doibase 10.1103/PhysRevLett.102.120502} {\bibfield
  {journal} {\bibinfo  {journal} {Phys. Rev. Lett.}\ }\textbf {\bibinfo
  {volume} {102}},\ \bibinfo {pages} {120502} (\bibinfo {year}
  {2009})}\BibitemShut {NoStop}%
\bibitem [{\citenamefont {West}\ \emph
  {et~al.}(2010{\natexlab{b}})\citenamefont {West}, \citenamefont {Fong},\ and\
  \citenamefont {Lidar}}]{West2010Near-Optimal}%
  \BibitemOpen
  \bibfield  {author} {\bibinfo {author} {\bibfnamefont {J.~R.}\ \bibnamefont
  {West}}, \bibinfo {author} {\bibfnamefont {B.~H.}\ \bibnamefont {Fong}}, \
  and\ \bibinfo {author} {\bibfnamefont {D.~A.}\ \bibnamefont {Lidar}},\ }\href
  {\doibase 10.1103/PhysRevLett.104.130501} {\bibfield  {journal} {\bibinfo
  {journal} {Phys. Rev. Lett.}\ }\textbf {\bibinfo {volume} {104}},\ \bibinfo
  {pages} {130501} (\bibinfo {year} {2010}{\natexlab{b}})}\BibitemShut
  {NoStop}%
\bibitem [{\citenamefont {Wang}\ \emph {et~al.}(2012)\citenamefont {Wang},
  \citenamefont {de~Lange}, \citenamefont {Rist\`e}, \citenamefont {Hanson},\
  and\ \citenamefont {Dobrovitski}}]{Wang2012Comparison}%
  \BibitemOpen
  \bibfield  {author} {\bibinfo {author} {\bibfnamefont {Z.-H.}\ \bibnamefont
  {Wang}}, \bibinfo {author} {\bibfnamefont {G.}~\bibnamefont {de~Lange}},
  \bibinfo {author} {\bibfnamefont {D.}~\bibnamefont {Rist\`e}}, \bibinfo
  {author} {\bibfnamefont {R.}~\bibnamefont {Hanson}}, \ and\ \bibinfo {author}
  {\bibfnamefont {V.~V.}\ \bibnamefont {Dobrovitski}},\ }\href {\doibase
  10.1103/PhysRevB.85.155204} {\bibfield  {journal} {\bibinfo  {journal} {Phys.
  Rev. B}\ }\textbf {\bibinfo {volume} {85}},\ \bibinfo {pages} {155204}
  (\bibinfo {year} {2012})}\BibitemShut {NoStop}%
\bibitem [{\citenamefont {Yannoni}\ and\ \citenamefont
  {Vieth}(1976)}]{Yannoni1976New}%
  \BibitemOpen
  \bibfield  {author} {\bibinfo {author} {\bibfnamefont {C.~S.}\ \bibnamefont
  {Yannoni}}\ and\ \bibinfo {author} {\bibfnamefont {H.-M.}\ \bibnamefont
  {Vieth}},\ }\href {\doibase 10.1103/PhysRevLett.37.1230} {\bibfield
  {journal} {\bibinfo  {journal} {Phys. Rev. Lett.}\ }\textbf {\bibinfo
  {volume} {37}},\ \bibinfo {pages} {1230} (\bibinfo {year}
  {1976})}\BibitemShut {NoStop}%
\bibitem [{\citenamefont {Zhang}\ \emph {et~al.}(2009)\citenamefont {Zhang},
  \citenamefont {Cappellaro}, \citenamefont {Antler}, \citenamefont {Pepper},
  \citenamefont {Cory}, \citenamefont {Dobrovitski}, \citenamefont
  {Ramanathan},\ and\ \citenamefont {Viola}}]{Zhang2009NMR}%
  \BibitemOpen
  \bibfield  {author} {\bibinfo {author} {\bibfnamefont {W.}~\bibnamefont
  {Zhang}}, \bibinfo {author} {\bibfnamefont {P.}~\bibnamefont {Cappellaro}},
  \bibinfo {author} {\bibfnamefont {N.}~\bibnamefont {Antler}}, \bibinfo
  {author} {\bibfnamefont {B.}~\bibnamefont {Pepper}}, \bibinfo {author}
  {\bibfnamefont {D.~G.}\ \bibnamefont {Cory}}, \bibinfo {author}
  {\bibfnamefont {V.~V.}\ \bibnamefont {Dobrovitski}}, \bibinfo {author}
  {\bibfnamefont {C.}~\bibnamefont {Ramanathan}}, \ and\ \bibinfo {author}
  {\bibfnamefont {L.}~\bibnamefont {Viola}},\ }\href {\doibase
  10.1103/PhysRevA.80.052323} {\bibfield  {journal} {\bibinfo  {journal} {Phys.
  Rev. A.}\ }\textbf {\bibinfo {volume} {80}},\ \bibinfo {pages} {052323}
  (\bibinfo {year} {2009})}\BibitemShut {NoStop}%
\bibitem [{\citenamefont {Gross}\ \emph {et~al.}(2008)\citenamefont {Gross},
  \citenamefont {Esteve}, \citenamefont {Giovanazzi}, \citenamefont {Weller},\
  and\ \citenamefont {Oberthaler}}]{Gross2008Squeezing}%
  \BibitemOpen
  \bibfield  {author} {\bibinfo {author} {\bibfnamefont {C.}~\bibnamefont
  {Gross}}, \bibinfo {author} {\bibfnamefont {J.}~\bibnamefont {Esteve}},
  \bibinfo {author} {\bibfnamefont {S.}~\bibnamefont {Giovanazzi}}, \bibinfo
  {author} {\bibfnamefont {A.}~\bibnamefont {Weller}}, \ and\ \bibinfo {author}
  {\bibfnamefont {M.}~\bibnamefont {Oberthaler}},\ }\href
  {https://doi.org/10.1038/nature07332} {\bibfield  {journal} {\bibinfo
  {journal} {Nature (London)}\ }\textbf {\bibinfo {volume} {455}},\ \bibinfo
  {pages} {1216} (\bibinfo {year} {2008})}\BibitemShut {NoStop}%
\bibitem [{\citenamefont {Degen}\ \emph {et~al.}(2017)\citenamefont {Degen},
  \citenamefont {Reinhard},\ and\ \citenamefont
  {Cappellaro}}]{Degen2017Quantum}%
  \BibitemOpen
  \bibfield  {author} {\bibinfo {author} {\bibfnamefont {C.~L.}\ \bibnamefont
  {Degen}}, \bibinfo {author} {\bibfnamefont {F.}~\bibnamefont {Reinhard}}, \
  and\ \bibinfo {author} {\bibfnamefont {P.}~\bibnamefont {Cappellaro}},\
  }\href {\doibase 10.1103/RevModPhys.89.035002} {\bibfield  {journal}
  {\bibinfo  {journal} {Rev. Mod. Phys.}\ }\textbf {\bibinfo {volume} {89}},\
  \bibinfo {pages} {035002} (\bibinfo {year} {2017})}\BibitemShut {NoStop}%
\bibitem [{\citenamefont {Luo}\ \emph {et~al.}(2017)\citenamefont {Luo},
  \citenamefont {Zou}, \citenamefont {Wu}, \citenamefont {Liu}, \citenamefont
  {Han}, \citenamefont {Tey},\ and\ \citenamefont
  {You}}]{Luo2017Deterministic}%
  \BibitemOpen
  \bibfield  {author} {\bibinfo {author} {\bibfnamefont {X.-Y.}\ \bibnamefont
  {Luo}}, \bibinfo {author} {\bibfnamefont {Y.-Q.}\ \bibnamefont {Zou}},
  \bibinfo {author} {\bibfnamefont {L.-N.}\ \bibnamefont {Wu}}, \bibinfo
  {author} {\bibfnamefont {Q.}~\bibnamefont {Liu}}, \bibinfo {author}
  {\bibfnamefont {M.-F.}\ \bibnamefont {Han}}, \bibinfo {author} {\bibfnamefont
  {M.~K.}\ \bibnamefont {Tey}}, \ and\ \bibinfo {author} {\bibfnamefont
  {L.}~\bibnamefont {You}},\ }\href
  {http://science.sciencemag.org/content/355/6325/620} {\bibfield  {journal}
  {\bibinfo  {journal} {Science}\ }\textbf {\bibinfo {volume} {355}},\ \bibinfo
  {pages} {620} (\bibinfo {year} {2017})}\BibitemShut {NoStop}%
\bibitem [{\citenamefont {Shenvi}\ \emph {et~al.}(2005)\citenamefont {Shenvi},
  \citenamefont {de~Sousa},\ and\ \citenamefont
  {Whaley}}]{Shenvi2005Universal}%
  \BibitemOpen
  \bibfield  {author} {\bibinfo {author} {\bibfnamefont {N.}~\bibnamefont
  {Shenvi}}, \bibinfo {author} {\bibfnamefont {R.}~\bibnamefont {de~Sousa}}, \
  and\ \bibinfo {author} {\bibfnamefont {K.~B.}\ \bibnamefont {Whaley}},\
  }\href {\doibase 10.1103/PhysRevB.71.224411} {\bibfield  {journal} {\bibinfo
  {journal} {Phys. Rev. B.}\ }\textbf {\bibinfo {volume} {71}},\ \bibinfo
  {pages} {224411} (\bibinfo {year} {2005})}\BibitemShut {NoStop}%
\bibitem [{\citenamefont {Merkulov}\ \emph {et~al.}(2002)\citenamefont
  {Merkulov}, \citenamefont {Efros},\ and\ \citenamefont
  {Rosen}}]{Merkulov2002Electron}%
  \BibitemOpen
  \bibfield  {author} {\bibinfo {author} {\bibfnamefont {I.~A.}\ \bibnamefont
  {Merkulov}}, \bibinfo {author} {\bibfnamefont {A.~L.}\ \bibnamefont {Efros}},
  \ and\ \bibinfo {author} {\bibfnamefont {M.}~\bibnamefont {Rosen}},\ }\href
  {\doibase 10.1103/PhysRevB.65.205309} {\bibfield  {journal} {\bibinfo
  {journal} {Phys. Rev. B.}\ }\textbf {\bibinfo {volume} {65}},\ \bibinfo
  {pages} {205309} (\bibinfo {year} {2002})}\BibitemShut {NoStop}%
\bibitem [{\citenamefont {Zhang}\ \emph {et~al.}(2016)\citenamefont {Zhang},
  \citenamefont {Han}, \citenamefont {Xu},\ and\ \citenamefont
  {Zhang}}]{zhang2016preserving}%
  \BibitemOpen
  \bibfield  {author} {\bibinfo {author} {\bibfnamefont {J.}~\bibnamefont
  {Zhang}}, \bibinfo {author} {\bibfnamefont {Y.}~\bibnamefont {Han}}, \bibinfo
  {author} {\bibfnamefont {P.}~\bibnamefont {Xu}}, \ and\ \bibinfo {author}
  {\bibfnamefont {W.}~\bibnamefont {Zhang}},\ }\href {\doibase
  10.1103/PhysRevA.94.053608} {\bibfield  {journal} {\bibinfo  {journal} {Phys.
  Rev. A}\ }\textbf {\bibinfo {volume} {94}},\ \bibinfo {pages} {053608}
  (\bibinfo {year} {2016})}\BibitemShut {NoStop}%
\bibitem [{\citenamefont {Stamper-Kurn}\ and\ \citenamefont
  {Ueda}(2013)}]{Stamper2013Spinor}%
  \BibitemOpen
  \bibfield  {author} {\bibinfo {author} {\bibfnamefont {D.~M.}\ \bibnamefont
  {Stamper-Kurn}}\ and\ \bibinfo {author} {\bibfnamefont {M.}~\bibnamefont
  {Ueda}},\ }\href {\doibase 10.1103/RevModPhys.85.1191} {\bibfield  {journal}
  {\bibinfo  {journal} {Rev. Mod. Phys.}\ }\textbf {\bibinfo {volume} {85}},\
  \bibinfo {pages} {1191} (\bibinfo {year} {2013})}\BibitemShut {NoStop}%
\bibitem [{\citenamefont {Law}\ \emph {et~al.}(1998)\citenamefont {Law},
  \citenamefont {Pu},\ and\ \citenamefont {Bigelow}}]{Law1998Quantum}%
  \BibitemOpen
  \bibfield  {author} {\bibinfo {author} {\bibfnamefont {C.~K.}\ \bibnamefont
  {Law}}, \bibinfo {author} {\bibfnamefont {H.}~\bibnamefont {Pu}}, \ and\
  \bibinfo {author} {\bibfnamefont {N.~P.}\ \bibnamefont {Bigelow}},\ }\href
  {\doibase 10.1103/PhysRevLett.81.5257} {\bibfield  {journal} {\bibinfo
  {journal} {Phys. Rev. Lett.}\ }\textbf {\bibinfo {volume} {81}},\ \bibinfo
  {pages} {5257} (\bibinfo {year} {1998})}\BibitemShut {NoStop}%
\bibitem [{\citenamefont {Yi}\ \emph {et~al.}(2002)\citenamefont {Yi},
  \citenamefont {M\"ustecapl\ifmmode \imath \else \i
  \fi{}o\ifmmode~\breve{g}\else \u{g}\fi{}lu}, \citenamefont {Sun},\ and\
  \citenamefont {You}}]{yi2002single}%
  \BibitemOpen
  \bibfield  {author} {\bibinfo {author} {\bibfnamefont {S.}~\bibnamefont
  {Yi}}, \bibinfo {author} {\bibfnamefont {O.~E.}\ \bibnamefont
  {M\"ustecapl\ifmmode \imath \else \i \fi{}o\ifmmode~\breve{g}\else
  \u{g}\fi{}lu}}, \bibinfo {author} {\bibfnamefont {C.~P.}\ \bibnamefont
  {Sun}}, \ and\ \bibinfo {author} {\bibfnamefont {L.}~\bibnamefont {You}},\
  }\href {\doibase 10.1103/PhysRevA.66.011601} {\bibfield  {journal} {\bibinfo
  {journal} {Phys. Rev. A}\ }\textbf {\bibinfo {volume} {66}},\ \bibinfo
  {pages} {011601} (\bibinfo {year} {2002})}\BibitemShut {NoStop}%
\bibitem [{\citenamefont {Eto}\ \emph {et~al.}(2014)\citenamefont {Eto},
  \citenamefont {Sadgrove}, \citenamefont {Hasegawa}, \citenamefont {Saito},\
  and\ \citenamefont {Hirano}}]{Eto2014Control}%
  \BibitemOpen
  \bibfield  {author} {\bibinfo {author} {\bibfnamefont {Y.}~\bibnamefont
  {Eto}}, \bibinfo {author} {\bibfnamefont {M.}~\bibnamefont {Sadgrove}},
  \bibinfo {author} {\bibfnamefont {S.}~\bibnamefont {Hasegawa}}, \bibinfo
  {author} {\bibfnamefont {H.}~\bibnamefont {Saito}}, \ and\ \bibinfo {author}
  {\bibfnamefont {T.}~\bibnamefont {Hirano}},\ }\href {\doibase
  10.1103/PhysRevA.90.013626} {\bibfield  {journal} {\bibinfo  {journal} {Phys.
  Rev. A.}\ }\textbf {\bibinfo {volume} {90}},\ \bibinfo {pages} {013626}
  (\bibinfo {year} {2014})}\BibitemShut {NoStop}%
\bibitem [{par()}]{parameternote}%
  \BibitemOpen
  \href@noop {} {}\bibinfo {note} {For a $^{87}$Rb spin-1 atomic condensate,
  the gyro magnetic ratio is $\gamma \approx 0.7$ MHz/G. In a typical stray
  magneitc field of 0.1 mG, the decoherence time from free evolution is about
  $T_{0.9} \approx 9.7$ ms. Our Uni-DD, under the magic condition with a bias
  field $B\approx 12.6$ mG and $\tau \approx 0.7$ ms, extends the spin
  coherence time for a CSS to about $T_{0.9} \approx 2.43$ s. While for a SSS,
  the decoherence time is extended from $T_{0.05} \approx 0.14$ ms in free
  evolution to $T_{0.05} \approx 42$ ms.}\BibitemShut {Stop}%
\bibitem [{\citenamefont {Kitagawa}\ and\ \citenamefont
  {Ueda}(1993)}]{Kitagawa1993Squeezed}%
  \BibitemOpen
  \bibfield  {author} {\bibinfo {author} {\bibfnamefont {M.}~\bibnamefont
  {Kitagawa}}\ and\ \bibinfo {author} {\bibfnamefont {M.}~\bibnamefont
  {Ueda}},\ }\href {\doibase 10.1103/PhysRevA.47.5138} {\bibfield  {journal}
  {\bibinfo  {journal} {Phys. Rev. A.}\ }\textbf {\bibinfo {volume} {47}},\
  \bibinfo {pages} {5138} (\bibinfo {year} {1993})}\BibitemShut {NoStop}%
\bibitem [{\citenamefont {Morsch}\ and\ \citenamefont
  {Oberthaler}(2006)}]{Morsch2006Dynamics}%
  \BibitemOpen
  \bibfield  {author} {\bibinfo {author} {\bibfnamefont {O.}~\bibnamefont
  {Morsch}}\ and\ \bibinfo {author} {\bibfnamefont {M.}~\bibnamefont
  {Oberthaler}},\ }\href {\doibase 10.1103/RevModPhys.78.179} {\bibfield
  {journal} {\bibinfo  {journal} {Rev. Mod. Phys.}\ }\textbf {\bibinfo {volume}
  {78}},\ \bibinfo {pages} {179} (\bibinfo {year} {2006})}\BibitemShut
  {NoStop}%
\bibitem [{\citenamefont {Jian}\ \emph {et~al.}(2011)\citenamefont {Jian},
  \citenamefont {Wang}, \citenamefont {Sun},\ and\ \citenamefont
  {Nori}}]{Jian2011Quantum}%
  \BibitemOpen
  \bibfield  {author} {\bibinfo {author} {\bibfnamefont {M.}~\bibnamefont
  {Jian}}, \bibinfo {author} {\bibfnamefont {X.}~\bibnamefont {Wang}}, \bibinfo
  {author} {\bibfnamefont {C.~P.}\ \bibnamefont {Sun}}, \ and\ \bibinfo
  {author} {\bibfnamefont {F.}~\bibnamefont {Nori}},\ }\href {\doibase
  https://doi.org/10.1016/j.physrep.2011.08.003} {\bibfield  {journal}
  {\bibinfo  {journal} {Phys. Rep.}\ }\textbf {\bibinfo {volume} {509}},\
  \bibinfo {pages} {89} (\bibinfo {year} {2011})}\BibitemShut {NoStop}%
\bibitem [{mcn()}]{mcnote}%
  \BibitemOpen
  \href@noop {} {}\bibinfo {note} {As shown in Figs. 2 and 3, the magic
  condition does not need to be strictly satisfied in practice, effective noise
  suppression is observed within a window of $2\gamma b_c$ centered at
  $\omega_m$ ($\gamma$ and $b_c$ are set to 1 in dimensionless units for the
  numerical results presented). The widths of the peaks ($\sim \gamma b_c$) in
  Figs. 2(b), 2(d), and 3(d) indicate that our Uni-DD is robust against
  deviations from the magic condition.}\BibitemShut {Stop}%
\bibitem [{\citenamefont {Fer}(1958)}]{Fer1958R}%
  \BibitemOpen
  \bibfield  {author} {\bibinfo {author} {\bibfnamefont {F.}~\bibnamefont
  {Fer}},\ }\href {https://doi.org/10.1063/1.1703993} {\bibfield  {journal}
  {\bibinfo  {journal} {Bull. Classe Sci. Acad. Roy. Belg.}\ }\textbf {\bibinfo
  {volume} {44}},\ \bibinfo {pages} {818} (\bibinfo {year} {1958})}\BibitemShut
  {NoStop}%
\bibitem [{\citenamefont {Takegoshi}\ \emph {et~al.}(2015)\citenamefont
  {Takegoshi}, \citenamefont {Miyazawa}, \citenamefont {Sharma},\ and\
  \citenamefont {Madhu}}]{takegoshi2015comparison}%
  \BibitemOpen
  \bibfield  {author} {\bibinfo {author} {\bibfnamefont {K.}~\bibnamefont
  {Takegoshi}}, \bibinfo {author} {\bibfnamefont {N.}~\bibnamefont {Miyazawa}},
  \bibinfo {author} {\bibfnamefont {K.}~\bibnamefont {Sharma}}, \ and\ \bibinfo
  {author} {\bibfnamefont {P.}~\bibnamefont {Madhu}},\ }\href {\doibase
  10.1063/1.4916324} {\bibfield  {journal} {\bibinfo  {journal} {J. Chem.
  Phys.}\ }\textbf {\bibinfo {volume} {142}},\ \bibinfo {pages} {134201}
  (\bibinfo {year} {2015})}\BibitemShut {NoStop}%
\bibitem [{\citenamefont {Loss}\ and\ \citenamefont
  {DiVincenzo}(1998)}]{Loss1998Quantum}%
  \BibitemOpen
  \bibfield  {author} {\bibinfo {author} {\bibfnamefont {D.}~\bibnamefont
  {Loss}}\ and\ \bibinfo {author} {\bibfnamefont {D.~P.}\ \bibnamefont
  {DiVincenzo}},\ }\href {\doibase 10.1103/PhysRevA.57.120} {\bibfield
  {journal} {\bibinfo  {journal} {Phys. Rev. A.}\ }\textbf {\bibinfo {volume}
  {57}},\ \bibinfo {pages} {120} (\bibinfo {year} {1998})}\BibitemShut
  {NoStop}%
\bibitem [{\citenamefont {Petta}\ \emph {et~al.}(2005)\citenamefont {Petta},
  \citenamefont {Johnson}, \citenamefont {Taylor}, \citenamefont {Laird},
  \citenamefont {Yacoby}, \citenamefont {Lukin}, \citenamefont {Marcus},
  \citenamefont {Hanson},\ and\ \citenamefont {Gossard}}]{Petta2005Coherent}%
  \BibitemOpen
  \bibfield  {author} {\bibinfo {author} {\bibfnamefont {J.~R.}\ \bibnamefont
  {Petta}}, \bibinfo {author} {\bibfnamefont {A.~C.}\ \bibnamefont {Johnson}},
  \bibinfo {author} {\bibfnamefont {J.~M.}\ \bibnamefont {Taylor}}, \bibinfo
  {author} {\bibfnamefont {E.~A.}\ \bibnamefont {Laird}}, \bibinfo {author}
  {\bibfnamefont {A.}~\bibnamefont {Yacoby}}, \bibinfo {author} {\bibfnamefont
  {M.~D.}\ \bibnamefont {Lukin}}, \bibinfo {author} {\bibfnamefont {C.~M.}\
  \bibnamefont {Marcus}}, \bibinfo {author} {\bibfnamefont {M.~P.}\
  \bibnamefont {Hanson}}, \ and\ \bibinfo {author} {\bibfnamefont {A.~C.}\
  \bibnamefont {Gossard}},\ }\href {\doibase 10.1126/science.1116955}
  {\bibfield  {journal} {\bibinfo  {journal} {Science}\ }\textbf {\bibinfo
  {volume} {309}},\ \bibinfo {pages} {2180} (\bibinfo {year}
  {2005})}\BibitemShut {NoStop}%
\bibitem [{\citenamefont {Taylor}\ \emph {et~al.}(2007)\citenamefont {Taylor},
  \citenamefont {Petta}, \citenamefont {Johnson}, \citenamefont {Yacoby},
  \citenamefont {Marcus},\ and\ \citenamefont {Lukin}}]{Taylor2007Relaxation}%
  \BibitemOpen
  \bibfield  {author} {\bibinfo {author} {\bibfnamefont {J.~M.}\ \bibnamefont
  {Taylor}}, \bibinfo {author} {\bibfnamefont {J.~R.}\ \bibnamefont {Petta}},
  \bibinfo {author} {\bibfnamefont {A.~C.}\ \bibnamefont {Johnson}}, \bibinfo
  {author} {\bibfnamefont {A.}~\bibnamefont {Yacoby}}, \bibinfo {author}
  {\bibfnamefont {C.~M.}\ \bibnamefont {Marcus}}, \ and\ \bibinfo {author}
  {\bibfnamefont {M.~D.}\ \bibnamefont {Lukin}},\ }\href {\doibase
  10.1103/PhysRevB.76.035315} {\bibfield  {journal} {\bibinfo  {journal} {Phys.
  Rev. B.}\ }\textbf {\bibinfo {volume} {76}},\ \bibinfo {pages} {035315}
  (\bibinfo {year} {2007})}\BibitemShut {NoStop}%
\bibitem [{\citenamefont {Paget}\ \emph {et~al.}(1977)\citenamefont {Paget},
  \citenamefont {Lampel}, \citenamefont {Sapoval},\ and\ \citenamefont
  {Safarov}}]{paget1977low}%
  \BibitemOpen
  \bibfield  {author} {\bibinfo {author} {\bibfnamefont {D.}~\bibnamefont
  {Paget}}, \bibinfo {author} {\bibfnamefont {G.}~\bibnamefont {Lampel}},
  \bibinfo {author} {\bibfnamefont {B.}~\bibnamefont {Sapoval}}, \ and\
  \bibinfo {author} {\bibfnamefont {V.~I.}\ \bibnamefont {Safarov}},\ }\href
  {\doibase 10.1103/PhysRevB.15.5780} {\bibfield  {journal} {\bibinfo
  {journal} {Phys. Rev. B}\ }\textbf {\bibinfo {volume} {15}},\ \bibinfo
  {pages} {5780} (\bibinfo {year} {1977})}\BibitemShut {NoStop}%
\bibitem [{\citenamefont {Dobrovitski}\ \emph {et~al.}(2006)\citenamefont
  {Dobrovitski}, \citenamefont {Taylor},\ and\ \citenamefont
  {Lukin}}]{dobrovitski2006long}%
  \BibitemOpen
  \bibfield  {author} {\bibinfo {author} {\bibfnamefont {V.~V.}\ \bibnamefont
  {Dobrovitski}}, \bibinfo {author} {\bibfnamefont {J.~M.}\ \bibnamefont
  {Taylor}}, \ and\ \bibinfo {author} {\bibfnamefont {M.~D.}\ \bibnamefont
  {Lukin}},\ }\href {\doibase 10.1103/PhysRevB.73.245318} {\bibfield  {journal}
  {\bibinfo  {journal} {Phys. Rev. B}\ }\textbf {\bibinfo {volume} {73}},\
  \bibinfo {pages} {245318} (\bibinfo {year} {2006})}\BibitemShut {NoStop}%
\bibitem [{\citenamefont {Zhang}\ \emph {et~al.}(2006)\citenamefont {Zhang},
  \citenamefont {Dobrovitski}, \citenamefont {Al-Hassanieh}, \citenamefont
  {Dagotto},\ and\ \citenamefont {Harmon}}]{zhang2006hyperfine}%
  \BibitemOpen
  \bibfield  {author} {\bibinfo {author} {\bibfnamefont {W.}~\bibnamefont
  {Zhang}}, \bibinfo {author} {\bibfnamefont {V.~V.}\ \bibnamefont
  {Dobrovitski}}, \bibinfo {author} {\bibfnamefont {K.~A.}\ \bibnamefont
  {Al-Hassanieh}}, \bibinfo {author} {\bibfnamefont {E.}~\bibnamefont
  {Dagotto}}, \ and\ \bibinfo {author} {\bibfnamefont {B.~N.}\ \bibnamefont
  {Harmon}},\ }\href {\doibase 10.1103/PhysRevB.74.205313} {\bibfield
  {journal} {\bibinfo  {journal} {Phys. Rev. B}\ }\textbf {\bibinfo {volume}
  {74}},\ \bibinfo {pages} {205313} (\bibinfo {year} {2006})}\BibitemShut
  {NoStop}%
\bibitem [{\citenamefont {Coish}\ \emph {et~al.}(2008)\citenamefont {Coish},
  \citenamefont {Fischer},\ and\ \citenamefont {Loss}}]{coish2008exponential}%
  \BibitemOpen
  \bibfield  {author} {\bibinfo {author} {\bibfnamefont {W.~A.}\ \bibnamefont
  {Coish}}, \bibinfo {author} {\bibfnamefont {J.}~\bibnamefont {Fischer}}, \
  and\ \bibinfo {author} {\bibfnamefont {D.}~\bibnamefont {Loss}},\ }\href
  {\doibase 10.1103/PhysRevB.77.125329} {\bibfield  {journal} {\bibinfo
  {journal} {Phys. Rev. B}\ }\textbf {\bibinfo {volume} {77}},\ \bibinfo
  {pages} {125329} (\bibinfo {year} {2008})}\BibitemShut {NoStop}%
\bibitem [{\citenamefont {Dobrovitski}\ and\ \citenamefont
  {De~Raedt}(2003)}]{dobrovitski2003efficient}%
  \BibitemOpen
  \bibfield  {author} {\bibinfo {author} {\bibfnamefont {V.~V.}\ \bibnamefont
  {Dobrovitski}}\ and\ \bibinfo {author} {\bibfnamefont {H.~A.}\ \bibnamefont
  {De~Raedt}},\ }\href {\doibase 10.1103/PhysRevE.67.056702} {\bibfield
  {journal} {\bibinfo  {journal} {Phys. Rev. E}\ }\textbf {\bibinfo {volume}
  {67}},\ \bibinfo {pages} {056702} (\bibinfo {year} {2003})}\BibitemShut
  {NoStop}%
\bibitem [{\citenamefont {Popescu}\ \emph {et~al.}(2006)\citenamefont
  {Popescu}, \citenamefont {Short},\ and\ \citenamefont
  {Winter}}]{Popescu2006Entanglement}%
  \BibitemOpen
  \bibfield  {author} {\bibinfo {author} {\bibfnamefont {S.}~\bibnamefont
  {Popescu}}, \bibinfo {author} {\bibfnamefont {A.~J.}\ \bibnamefont {Short}},
  \ and\ \bibinfo {author} {\bibfnamefont {A.}~\bibnamefont {Winter}},\ }\href
  {https://doi.org/10.1038/nphys444} {\bibfield  {journal} {\bibinfo  {journal}
  {Nat. Phys.}\ }\textbf {\bibinfo {volume} {2}},\ \bibinfo {pages} {754}
  (\bibinfo {year} {2006})}\BibitemShut {NoStop}%
\bibitem [{bia()}]{biasnote}%
  \BibitemOpen
  \href@noop {} {}\bibinfo {note} {The residual uncompensated term to leading
  order in the biaxial PDD is $\sim \gamma h_{x,y,z} S_{x,y,z}\tau \propto
  \tau^1$, which is the same as in the Uni-DD we propose, since
  $S_y(h_zh_y+h_yh_z)/(2\omega) \propto \tau^1$ under the magic condition
  $\omega \tau = 2\pi$. Therefore, our Uni-DD exhibits a comparable level of
  noise suppression to the PDD, although the number of pulses required is
  halved. The coherence time is extended by roughly $\omega/h_z$ times that of
  the decoherence time under free evolution without DD. For the Uni-DD to be
  effective, the bias field $B=\omega/\gamma $ needs to be 1-2 orders of
  magnitude larger than the stray fields. We have observed that in general the
  larger the bias field, the better the performance of the Uni-DD.}\BibitemShut
  {Stop}%
\bibitem [{\citenamefont {Carr}\ and\ \citenamefont
  {Purcell}(1954)}]{Carr1954Effects}%
  \BibitemOpen
  \bibfield  {author} {\bibinfo {author} {\bibfnamefont {H.~Y.}\ \bibnamefont
  {Carr}}\ and\ \bibinfo {author} {\bibfnamefont {E.~M.}\ \bibnamefont
  {Purcell}},\ }\href {\doibase 10.1103/PhysRev.94.630} {\bibfield  {journal}
  {\bibinfo  {journal} {Phys. Rev.}\ }\textbf {\bibinfo {volume} {94}},\
  \bibinfo {pages} {630} (\bibinfo {year} {1954})}\BibitemShut {NoStop}%
\bibitem [{\citenamefont {Meiboom}\ and\ \citenamefont
  {Gill}(1958)}]{Meiboom1958Modified}%
  \BibitemOpen
  \bibfield  {author} {\bibinfo {author} {\bibfnamefont {S.}~\bibnamefont
  {Meiboom}}\ and\ \bibinfo {author} {\bibfnamefont {D.}~\bibnamefont {Gill}},\
  }\href {https://doi.org/10.1063/1.1716296} {\bibfield  {journal} {\bibinfo
  {journal} {Rev. Sci. Instrum.}\ }\textbf {\bibinfo {volume} {29}},\ \bibinfo
  {pages} {688} (\bibinfo {year} {1958})}\BibitemShut {NoStop}%
\bibitem [{nvc()}]{nvcnote}%
  \BibitemOpen
  \href@noop {} {}\bibinfo {note} {The Uni-DD protocol significantly suppresses
  typical noises in a nitrogen vacancy (NV) center from dipolar coupling
  between the NV center spin and the surrounding nuclear spins, as shown in the
  SM.}\BibitemShut {Stop}%
\end{thebibliography}

\begin{thebibliography}{17}
\expandafter\ifx\csname natexlab\endcsname\relax\def\natexlab#1{#1}\fi
\expandafter\ifx\csname bibnamefont\endcsname\relax
  \def\bibnamefont#1{#1}\fi
\expandafter\ifx\csname bibfnamefont\endcsname\relax
  \def\bibfnamefont#1{#1}\fi
\expandafter\ifx\csname citenamefont\endcsname\relax
  \def\citenamefont#1{#1}\fi
\expandafter\ifx\csname url\endcsname\relax
  \def\url#1{\texttt{#1}}\fi
\expandafter\ifx\csname urlprefix\endcsname\relax\def\urlprefix{URL }\fi
\providecommand{\bibinfo}[2]{#2}
\providecommand{\eprint}[2][]{\url{#2}}

\bibitem[{\citenamefont{Ho}(1998)}]{Ho1998Spinor}
\bibinfo{author}{\bibfnamefont{T.-L.} \bibnamefont{Ho}},
  \bibinfo{journal}{Phys. Rev. Lett.} \textbf{\bibinfo{volume}{81}},
  \bibinfo{pages}{742} (\bibinfo{year}{1998}).

\bibitem[{\citenamefont{Ohmi and Machida}(1998)}]{Ohmi1998Bose}
\bibinfo{author}{\bibfnamefont{T.}~\bibnamefont{Ohmi}} \bibnamefont{and}
  \bibinfo{author}{\bibfnamefont{K.}~\bibnamefont{Machida}},
  \bibinfo{journal}{J. Phys. Soc. Jpn.} \textbf{\bibinfo{volume}{67}},
  \bibinfo{pages}{1822} (\bibinfo{year}{1998}).

\bibitem[{\citenamefont{Law et~al.}(1998)\citenamefont{Law, Pu, and
  Bigelow}}]{Law1998Quantum1}
\bibinfo{author}{\bibfnamefont{C.~K.} \bibnamefont{Law}},
  \bibinfo{author}{\bibfnamefont{H.}~\bibnamefont{Pu}}, \bibnamefont{and}
  \bibinfo{author}{\bibfnamefont{N.~P.} \bibnamefont{Bigelow}},
  \bibinfo{journal}{Phys. Rev. Lett.} \textbf{\bibinfo{volume}{81}},
  \bibinfo{pages}{5257} (\bibinfo{year}{1998}).

\bibitem[{\citenamefont{Zhang et~al.}(2003)\citenamefont{Zhang, Yi, and
  You}}]{Zhang2003Mean1}
\bibinfo{author}{\bibfnamefont{W.}~\bibnamefont{Zhang}},
  \bibinfo{author}{\bibfnamefont{S.}~\bibnamefont{Yi}}, \bibnamefont{and}
  \bibinfo{author}{\bibfnamefont{L.}~\bibnamefont{You}}, \bibinfo{journal}{New
  J. Phys.} \textbf{\bibinfo{volume}{5}}, \bibinfo{pages}{77}
  (\bibinfo{year}{2003}).

\bibitem[{\citenamefont{Zhang et~al.}(2016)\citenamefont{Zhang, Han, Xu, and
  Zhang}}]{zhang2016preserving1}
\bibinfo{author}{\bibfnamefont{J.}~\bibnamefont{Zhang}},
  \bibinfo{author}{\bibfnamefont{Y.}~\bibnamefont{Han}},
  \bibinfo{author}{\bibfnamefont{P.}~\bibnamefont{Xu}}, \bibnamefont{and}
  \bibinfo{author}{\bibfnamefont{W.}~\bibnamefont{Zhang}},
  \bibinfo{journal}{Phys. Rev. A.} \textbf{\bibinfo{volume}{94}},
  \bibinfo{pages}{053608} (\bibinfo{year}{2016}).

\bibitem[{\citenamefont{Fer}(1958)}]{Fer1958R1}
\bibinfo{author}{\bibfnamefont{F.}~\bibnamefont{Fer}}, \bibinfo{journal}{Bull.
  Classe Sci. Acad. Roy. Belg.} \textbf{\bibinfo{volume}{44}},
  \bibinfo{pages}{818} (\bibinfo{year}{1958}).

\bibitem[{\citenamefont{Takegoshi et~al.}(2015)\citenamefont{Takegoshi,
  Miyazawa, Sharma, and Madhu}}]{takegoshi2015comparison1}
\bibinfo{author}{\bibfnamefont{K.}~\bibnamefont{Takegoshi}},
  \bibinfo{author}{\bibfnamefont{N.}~\bibnamefont{Miyazawa}},
  \bibinfo{author}{\bibfnamefont{K.}~\bibnamefont{Sharma}}, \bibnamefont{and}
  \bibinfo{author}{\bibfnamefont{P.}~\bibnamefont{Madhu}}, \bibinfo{journal}{J.
  Chem. Phys.} \textbf{\bibinfo{volume}{142}}, \bibinfo{pages}{134201}
  (\bibinfo{year}{2015}).

\bibitem[{\citenamefont{Haeberlen and Waugh}(1968)}]{Haeberlen1968Coherent}
\bibinfo{author}{\bibfnamefont{U.}~\bibnamefont{Haeberlen}} \bibnamefont{and}
  \bibinfo{author}{\bibfnamefont{J.~S.} \bibnamefont{Waugh}},
  \bibinfo{journal}{Phys. Rev.} \textbf{\bibinfo{volume}{175}},
  \bibinfo{pages}{453} (\bibinfo{year}{1968}).

\bibitem[{\citenamefont{Haeberlen}(1976)}]{haeberlen1976high}
\bibinfo{author}{\bibfnamefont{U.}~\bibnamefont{Haeberlen}},
  \emph{\bibinfo{title}{High Resolution NMR in solids: selective averaging}}
  (\bibinfo{publisher}{Academic, New York}, \bibinfo{year}{1976}).

\bibitem[{\citenamefont{Mehring}(1983)}]{mehring1983principles}
\bibinfo{author}{\bibfnamefont{M.}~\bibnamefont{Mehring}},
  \emph{\bibinfo{title}{Principles of high resolution NMR in solids}}
  (\bibinfo{publisher}{Springer-Verleg, Berlin}, \bibinfo{year}{1983}),
  \bibinfo{edition}{2nd} ed.

\bibitem[{\citenamefont{Loss and DiVincenzo}(1998)}]{Loss1998Quantum1}
\bibinfo{author}{\bibfnamefont{D.}~\bibnamefont{Loss}} \bibnamefont{and}
  \bibinfo{author}{\bibfnamefont{D.~P.} \bibnamefont{DiVincenzo}},
  \bibinfo{journal}{Phys. Rev. A.} \textbf{\bibinfo{volume}{57}},
  \bibinfo{pages}{120} (\bibinfo{year}{1998}).

\bibitem[{\citenamefont{Petta et~al.}(2005)\citenamefont{Petta, Johnson,
  Taylor, Laird, Yacoby, Lukin, Marcus, Hanson, and
  Gossard}}]{Petta2005Coherent1}
\bibinfo{author}{\bibfnamefont{J.~R.} \bibnamefont{Petta}},
  \bibinfo{author}{\bibfnamefont{A.~C.} \bibnamefont{Johnson}},
  \bibinfo{author}{\bibfnamefont{J.~M.} \bibnamefont{Taylor}},
  \bibinfo{author}{\bibfnamefont{E.~A.} \bibnamefont{Laird}},
  \bibinfo{author}{\bibfnamefont{A.}~\bibnamefont{Yacoby}},
  \bibinfo{author}{\bibfnamefont{M.~D.} \bibnamefont{Lukin}},
  \bibinfo{author}{\bibfnamefont{C.~M.} \bibnamefont{Marcus}},
  \bibinfo{author}{\bibfnamefont{M.~P.} \bibnamefont{Hanson}},
  \bibnamefont{and} \bibinfo{author}{\bibfnamefont{A.~C.}
  \bibnamefont{Gossard}}, \bibinfo{journal}{Science}
  \textbf{\bibinfo{volume}{309}}, \bibinfo{pages}{2180} (\bibinfo{year}{2005}).

\bibitem[{\citenamefont{Koppens et~al.}(2006)\citenamefont{Koppens, Buizert,
  Tielrooij, Vink, Nowack, Meunier, Kouwenhoven, and
  Vandersypen}}]{Koppens2006Driven1}
\bibinfo{author}{\bibfnamefont{F.~H.} \bibnamefont{Koppens}},
  \bibinfo{author}{\bibfnamefont{C.}~\bibnamefont{Buizert}},
  \bibinfo{author}{\bibfnamefont{K.~J.} \bibnamefont{Tielrooij}},
  \bibinfo{author}{\bibfnamefont{I.~T.} \bibnamefont{Vink}},
  \bibinfo{author}{\bibfnamefont{K.~C.} \bibnamefont{Nowack}},
  \bibinfo{author}{\bibfnamefont{T.}~\bibnamefont{Meunier}},
  \bibinfo{author}{\bibfnamefont{L.~P.} \bibnamefont{Kouwenhoven}},
  \bibnamefont{and} \bibinfo{author}{\bibfnamefont{L.~M.}
  \bibnamefont{Vandersypen}}, \bibinfo{journal}{Nature (London)}
  \textbf{\bibinfo{volume}{442}}, \bibinfo{pages}{766} (\bibinfo{year}{2006}).

\bibitem[{\citenamefont{Taylor et~al.}(2007)\citenamefont{Taylor, Petta,
  Johnson, Yacoby, Marcus, and Lukin}}]{Taylor2007Relaxation1}
\bibinfo{author}{\bibfnamefont{J.~M.} \bibnamefont{Taylor}},
  \bibinfo{author}{\bibfnamefont{J.~R.} \bibnamefont{Petta}},
  \bibinfo{author}{\bibfnamefont{A.~C.} \bibnamefont{Johnson}},
  \bibinfo{author}{\bibfnamefont{A.}~\bibnamefont{Yacoby}},
  \bibinfo{author}{\bibfnamefont{C.~M.} \bibnamefont{Marcus}},
  \bibnamefont{and} \bibinfo{author}{\bibfnamefont{M.~D.} \bibnamefont{Lukin}},
  \bibinfo{journal}{Phys. Rev. B.} \textbf{\bibinfo{volume}{76}},
  \bibinfo{pages}{035315} (\bibinfo{year}{2007}).

\bibitem[{\citenamefont{Al-Hassanieh et~al.}(2006)\citenamefont{Al-Hassanieh,
  Dobrovitski, Dagotto, and Harmon}}]{al2006numerical}
\bibinfo{author}{\bibfnamefont{K.~A.} \bibnamefont{Al-Hassanieh}},
  \bibinfo{author}{\bibfnamefont{V.}~\bibnamefont{Dobrovitski}},
  \bibinfo{author}{\bibfnamefont{E.}~\bibnamefont{Dagotto}}, \bibnamefont{and}
  \bibinfo{author}{\bibfnamefont{B.}~\bibnamefont{Harmon}},
  \bibinfo{journal}{Phys. Rev. Lett.} \textbf{\bibinfo{volume}{97}},
  \bibinfo{pages}{037204} (\bibinfo{year}{2006}).

\bibitem[{\citenamefont{Zhang et~al.}(2006)\citenamefont{Zhang, Dobrovitski,
  Al-Hassanieh, Dagotto, and Harmon}}]{zhang2006hyperfine1}
\bibinfo{author}{\bibfnamefont{W.}~\bibnamefont{Zhang}},
  \bibinfo{author}{\bibfnamefont{V.}~\bibnamefont{Dobrovitski}},
  \bibinfo{author}{\bibfnamefont{K.~A.} \bibnamefont{Al-Hassanieh}},
  \bibinfo{author}{\bibfnamefont{E.}~\bibnamefont{Dagotto}}, \bibnamefont{and}
  \bibinfo{author}{\bibfnamefont{B.}~\bibnamefont{Harmon}},
  \bibinfo{journal}{Phys. Rev. B.} \textbf{\bibinfo{volume}{74}},
  \bibinfo{pages}{205313} (\bibinfo{year}{2006}).

\bibitem[{\citenamefont{Zhang et~al.}(2007{\natexlab{a}})\citenamefont{Zhang,
  Dobrovitski, Santos, Viola, and Harmon}}]{zhang2007dynamical1}
\bibinfo{author}{\bibfnamefont{W.}~\bibnamefont{Zhang}},
  \bibinfo{author}{\bibfnamefont{V.}~\bibnamefont{Dobrovitski}},
  \bibinfo{author}{\bibfnamefont{L.~F.} \bibnamefont{Santos}},
  \bibinfo{author}{\bibfnamefont{L.}~\bibnamefont{Viola}}, \bibnamefont{and}
  \bibinfo{author}{\bibfnamefont{B.}~\bibnamefont{Harmon}},
  \bibinfo{journal}{Phys. Rev. B.} \textbf{\bibinfo{volume}{75}},
  \bibinfo{pages}{201302} (\bibinfo{year}{2007}{\natexlab{a}}).

\bibitem[{\citenamefont{Khodjasteh and Lidar}(2005)}]{khodjasteh2005Fault}
\bibinfo{author}{\bibfnamefont{K.}~\bibnamefont{Khodjasteh}} \bibnamefont{and}
  \bibinfo{author}{\bibfnamefont{D.~A.} \bibnamefont{Lidar}},
  \bibinfo{journal}{Phys. Rev. Lett.} \textbf{\bibinfo{volume}{95}},
  \bibinfo{pages}{180501} (\bibinfo{year}{2005}).

\bibitem[{\citenamefont{Zhang et~al.}(2007{\natexlab{b}})\citenamefont{Zhang,
  Dobrovitski, Santos, Viola, and Harmon}}]{zhang2007suppression}
\bibinfo{author}{\bibfnamefont{W.}~\bibnamefont{Zhang}},
  \bibinfo{author}{\bibfnamefont{V.}~\bibnamefont{Dobrovitski}},
  \bibinfo{author}{\bibfnamefont{L.~F.} \bibnamefont{Santos}},
  \bibinfo{author}{\bibfnamefont{L.}~\bibnamefont{Viola}}, \bibnamefont{and}
  \bibinfo{author}{\bibfnamefont{B.}~\bibnamefont{Harmon}},
  \bibinfo{journal}{J. Mod. Opt.} \textbf{\bibinfo{volume}{54}},
  \bibinfo{pages}{2629} (\bibinfo{year}{2007}{\natexlab{b}}).

\bibitem[{\citenamefont{Zhang et~al.}(2008)\citenamefont{Zhang, Konstantinidis,
  Dobrovitski, Harmon, Santos, and Viola}}]{Zhang2008Long1}
\bibinfo{author}{\bibfnamefont{W.}~\bibnamefont{Zhang}},
  \bibinfo{author}{\bibfnamefont{N.~P.} \bibnamefont{Konstantinidis}},
  \bibinfo{author}{\bibfnamefont{V.~V.} \bibnamefont{Dobrovitski}},
  \bibinfo{author}{\bibfnamefont{B.~N.} \bibnamefont{Harmon}},
  \bibinfo{author}{\bibfnamefont{L.~F.} \bibnamefont{Santos}},
  \bibnamefont{and} \bibinfo{author}{\bibfnamefont{L.}~\bibnamefont{Viola}},
  \bibinfo{journal}{Phys. Rev. B.} \textbf{\bibinfo{volume}{77}},
  \bibinfo{pages}{125336} (\bibinfo{year}{2008}).

\bibitem[{\citenamefont{Uhrig}(2007)}]{Uhrig2007keeping1}
\bibinfo{author}{\bibfnamefont{G.~S.} \bibnamefont{Uhrig}},
  \bibinfo{journal}{Phys. Rev. Lett.} \textbf{\bibinfo{volume}{98}},
  \bibinfo{pages}{100504} (\bibinfo{year}{2007}).

\bibitem[{\citenamefont{Uhrig}(2009)}]{Uhrig2009Concatenated1}
\bibinfo{author}{\bibfnamefont{G.~S.} \bibnamefont{Uhrig}},
  \bibinfo{journal}{Phys. Rev. Lett.} \textbf{\bibinfo{volume}{102}},
  \bibinfo{pages}{120502} (\bibinfo{year}{2009}).

\bibitem[{\citenamefont{Wang et~al.}(2012)\citenamefont{Wang, de~Lange,
  Rist\`e, Hanson, and Dobrovitski}}]{Wang2012Comparison1}
\bibinfo{author}{\bibfnamefont{Z.-H.} \bibnamefont{Wang}},
  \bibinfo{author}{\bibfnamefont{G.}~\bibnamefont{de~Lange}},
  \bibinfo{author}{\bibfnamefont{D.}~\bibnamefont{Rist\`e}},
  \bibinfo{author}{\bibfnamefont{R.}~\bibnamefont{Hanson}}, \bibnamefont{and}
  \bibinfo{author}{\bibfnamefont{V.~V.} \bibnamefont{Dobrovitski}},
  \bibinfo{journal}{Phys. Rev. B} \textbf{\bibinfo{volume}{85}},
  \bibinfo{pages}{155204} (\bibinfo{year}{2012}).

\bibitem[{\citenamefont{West et~al.}(2010)\citenamefont{West, Fong, and
  Lidar}}]{West2010Near-Optimal1}
\bibinfo{author}{\bibfnamefont{J.~R.} \bibnamefont{West}},
  \bibinfo{author}{\bibfnamefont{B.~H.} \bibnamefont{Fong}}, \bibnamefont{and}
  \bibinfo{author}{\bibfnamefont{D.~A.} \bibnamefont{Lidar}},
  \bibinfo{journal}{Phys. Rev. Lett.} \textbf{\bibinfo{volume}{104}},
  \bibinfo{pages}{130501} (\bibinfo{year}{2010}).

\bibitem[{\citenamefont{Hall et~al.}(2014)\citenamefont{Hall, Cole, and
  Hollenberg}}]{Hall2014PRB}
\bibinfo{author}{\bibfnamefont{L.~T.} \bibnamefont{Hall}},
  \bibinfo{author}{\bibfnamefont{J.~H.} \bibnamefont{Cole}}, \bibnamefont{and}
  \bibinfo{author}{\bibfnamefont{L.~C.~L.} \bibnamefont{Hollenberg}},
  \bibinfo{journal}{Phys. Rev. B} \textbf{\bibinfo{volume}{90}},
  \bibinfo{pages}{075201} (\bibinfo{year}{2014}).

\end{thebibliography}
\end{document}